\documentclass[11pt]{article}
\usepackage{geometry}                
\usepackage{graphicx}
\usepackage{amssymb}
\usepackage{amsmath}
\usepackage{ulem}
\usepackage{epstopdf}
\def\vp{^{\vphantom{.}}}

\title{The Ising susceptibility scaling function}
\author{Y. Chan%
\footnote{Department of Mathematics and Statistics, The University of Melbourne, Parkville, Victoria 3052, Australia.},
A. J. Guttmann,$\!{}^{\mbox{\scriptsize\thefootnote}}$
B. G. Nickel%
\footnote{Department of Physics, University of Guelph, Guelph, Ontario, Canada N1G 2W1.}\ \ and
J. H. H. Perk%
\footnote{Department of Physics, Oklahoma State University, Stillwater, Oklahoma 74078-3072, USA.}
\footnote{Department of Theoretical Physics, (RSPE), and Centre for Mathematics and its Applications (CMA), Australian National University, Canberra, ACT 2600, Australia.}}

\begin{document}
\maketitle
\begin{abstract}
We have dramatically extended the zero field susceptibility series at both high and low temperature of the Ising model on the triangular and honeycomb lattices, and used these data and newly available further terms for the square lattice to calculate a number of terms in the scaling function expansion around both the ferromagnetic and, for the square and honeycomb lattices, the antiferromagnetic critical point.

\end{abstract}

Cyril Domb was a pioneer in the application of series expansions to the study of critical phenomena \cite{Domb1,Domb2}. He encouraged many colleagues to develop this approach and headed a group, the ``Kings College group," who applied his ideas to investigate the behaviour of co-operative assemblies and percolation processes with considerable success. Domb's unselfish and generous attitude in urging people to follow up and develop the series approach was an important factor in the subsequent evolution of research in these areas. It is therefore with considerable pleasure that we dedicate this paper to Cyril Domb, on the occasion of his 90th birthday. In it we show just how powerful the series approach can be, as we present an analysis based on hundreds, and in some cases thousands of terms in the expansion of the susceptibility of the two-dimensional Ising model. It would be fair to say that no method other than the series method provides anything remotely approaching this level of information about the susceptibility.

\section{Introduction\label{sec:intro}}
A decade ago a number of the current authors reported on a substantial extension of the square lattice Ising susceptibility series to some 300 terms \cite{ONGP2,ONGP}. We found breakdown of the simple scaling picture that assumes the absence of irrelevant scaling fields. The first breakdown,
which was identified with the breakdown of rotational symmetry of the square lattice, occurred at O$(\tau^4)$, with $\tau$ to leading order proportional to the temperature deviation from critical, $T-T_{\mathrm{c}}$. A second breakdown was identified at O$(\tau^6),$ ascribed to an additional irrelevant variable. At the time it was foreshadowed that the corresponding calculation for the triangular and honeycomb lattices would be necessary in order to distinguish between lattice effects and more fundamental breakdowns intrinsic to the model.

In this study we report on the derivation and analysis of triangular and honeycomb lattice series to more than 300 terms, followed by a calculation of the corresponding scaling functions. Our numerical work is of sufficient accuracy that we can unambiguously identify the same irrational constant, that appeared at O$(\tau^6)$ in the square lattice scaling function and was ascribed to a second irrelevant variable, as a contribution to O$(\tau^6)$ in both triangular and honeycomb lattices. Furthermore, we find another irrational constant common to all lattices at O$(\tau^{10})$ which can be ascribed to yet another (third) irrelevant variable. These results clearly indicate aspects of universality in the susceptibility beyond those found at leading order. 

A limited selection of our results which are the basis for these remarks on universality are given in the immediately following text while the very extensive complete listing can be found in the appendices. In subsequent sections we elaborate on the results below and give details of how they were obtained. Specifically, in section \ref{sec:scaling} we put our results in the context of scaling theory and speculate on the identification of our correction to scaling terms with the operators of the conformal field theory that describes the Ising model. Section \ref{sec:series} describes how the series expansions were obtained from the quadratic recurrence relations for the $Z$-invariant Ising model specialized to the triangular/honeycomb system. In section \ref{sec:extract} we describe some of the series analysis details, in particular those aspects that differ from what was done in \cite{ONGP}. 

Our numerical work indicates that the reduced susceptibility on any lattice near the ferromagnetic critical point (for $T>T_{\mathrm{c}}$ or $T<T_{\mathrm{c}}$) is given by\footnote{The notation here differs from that in \cite{ONGP} and the earlier literature in that for a common treatment of all lattices it is convenient to absorb a factor $(2K_{\mathrm{c}}\sqrt{2})^{7/4}$ into the definition of $C_{0\pm}$, cf.\ equation (\ref{5}) vs.\ the appendix in \cite{ONGP}.}
\begin{equation}\label{chiform}
\bar \chi_\pm^{lattice}\equiv
k_{\mathrm{B}}T\chi_\pm^{lattice}=C_{0\pm}^{lattice}|\tau|^{-7/4}F_\pm^{lattice}+B^{lattice},
\end{equation}
where $B$ is the contribution of the ``short-distance" terms and includes an analytic background. It is of the form
\begin{equation}\label{background}
B = \sum_{q=0}^\infty \sum_{p=0}^{\lfloor \sqrt{q} \rfloor} b^{(p,q)}(\log |\tau| )^p\tau^q
\end{equation}
with the $b^{(p,q)}$ the same above and below $T_{\mathrm{c}}$ but, of course, different for each lattice. The temperature variable $\tau$ is simply related to the low-temperature elliptic parameter $k$ ($\equiv k_<$) by the same expressions 
\begin{equation}\label{3}
\tau=\frac{1}{2}\left(\sqrt{k}-\frac{1}{\sqrt{k}}\right),\hspace{1cm}k =(\tau+\sqrt{1+\tau^2})^2
\end{equation}
for every lattice. The elliptic parameter $k$ depends on the lattice; we have, with $K=J/k_{\mathrm{B}}T$,
\begin{eqnarray}
&k_{\mathrm{sq}} = 1/s^2,\quad s = \sinh{2K_{\mathrm{sq}}},\quad {\rm square},\nonumber\\\cr
&\displaystyle
k_{\mathrm{tr}} = \frac{4u^{3/2}}{(1-u)^{3/2}\sqrt{1+3u}},\quad u = \exp{(-4K_{\mathrm{tr}})},\quad {\rm triangular},\nonumber\\\cr
&\displaystyle
k_{\mathrm{hc}} = \frac{4z^{3/2}\sqrt{1-z+z^2}}{(1-z)^3(1+z)},\quad z = \exp{(-2K_{\mathrm{hc}})},\quad {\rm honeycomb}.\label{k}
\end{eqnarray}
Duality relates the high-temperature elliptic parameter $k_>$ to the low-temperature one by $k_> = 1/k_<$ or, what is equivalent, by the replacement $\tau \rightarrow - \tau$. Furthermore, since the honeycomb lattice is the dual of the triangular lattice and is also related by a star-triangle transformation, we can take $k_{\mathrm{tr}}=k_{\mathrm{hc}}$ as a common elliptic parameter $k_<$ with the $u$(triangle) and $z$(honeycomb) then connected by
\begin{equation}\label{star-triangle}
u = \frac{z}{1-z+z^2}, \hspace{1cm} z = \frac{2u}{1+u+\sqrt{(1-u)(1+3u)}}.
\end{equation}

The $C_{0\pm}$ constants in (\ref{chiform}) for the different lattices are
related as follows. First, we define $C_{0\pm}$ as the values for
the square lattice, that is\footnote{For the calculation of $C_{0\pm}$ see the footnote on page 3904 in \cite{Nic1}. Here we have used 
predictor-correctors of order as high as 25. This approach uses the Painlev\'e III equation of \cite{Wu76,BMW73,TM73}. Alternatively, one can also use the Painlev\'e V formulation \cite{JM17,AP2}.}
\begin{eqnarray}\label{5}
&&C_{0+} \equiv C_{0+}^{\mathrm{sq}} = 1.00081526044021264711947636304721023693753492559778\backslash\nonumber\\
&&\hspace{6em}92751083189882604491051665192385157187485052515870678\,\sqrt{2}, \\
&&C_{0-} \equiv C_{0-}^{\mathrm{sq}} = 1.0009603287252621894809349551720973205725059517701173\backslash\nonumber\\
&&\hspace{6em}61531948595158755619871466228353934981038826872108\,\sqrt{2}/(12\pi).\nonumber
\end{eqnarray}
Then
\begin{equation}\label{6}
C_{0\pm}^{\mathrm{tr}} = 4 C_{0\pm}/\sqrt{27}, \hspace{1cm} C_{0\pm}^{\mathrm{hc}} = 8 C_{0\pm}/\sqrt{27},
\end{equation}
as follows from lattice-lattice scaling \cite{Gu74,RB75} or $Z$-invariance \cite{AP03}.
The scaling functions through O($\tau^{10}$) are
\begin{eqnarray}
\label{fnew}
F_{\pm}^{\mathrm{sq}} & = & k^{1/4} \left[ 1 + \frac{\tau^2}{2} - \frac{\tau^4}{12} + \left(\frac{647}{15360} - \frac{7 C_{6\pm}}{5}\right) \tau^6 - \left( \frac{296813}{11059200} - \frac{4973 C_{6\pm}}{3600} \right) \tau^8 \right. \nonumber \\
	& & \hspace{1cm} + \left. \left( \frac{23723921}{1238630400} - \frac{100261 C_{6\pm}}{115200} - \frac{793 C_{10\pm}}{210} \right) \tau^{10} \right], \nonumber \\
F_{\pm}^{\mathrm{tr}} & = & k^{1/4} \left[ 1 + \frac{\tau^2}{2} - \frac{21\tau^4}{256} + \left(\frac{85}{2048} - \frac{3 C_{6\pm}}{2} \right) \tau^6 - \left(\frac{43361}{1638400} - \frac{1209 C_{6\pm}}{800} \right) \tau^8 \right. \nonumber \\
	& & \hspace{1cm} + \left. \left( \frac{1734121}{91750400} - \frac{261 C_{6\pm}}{200} - \frac{51 C_{10\pm}}{70} \right) \tau^{10} \right], \\
F_{\pm}^{\mathrm{hc}} & = & k^{1/4} \left[ 1 + \frac{\tau^2}{2} - \frac{21\tau^4}{256} + \left(\frac{85}{2048} - \frac{C_{6\pm}}{2} \right) \tau^6 - \left( \frac{43361}{1638400} - \frac{409 C_{6\pm}}{800} \right) \tau^8 \right. \nonumber \\
	& & \hspace{1cm} + \left. \left( \frac{1734121}{91750400} - \frac{61 C_{6\pm}}{200} - \frac{121 C_{10\pm}}{70} \right) \tau^{10} \right],\nonumber
\end{eqnarray}
where
\begin{eqnarray}\label{8}
&&C_{6-\phantom{1}}=
4.54530659737804996885745146127924976519048127125911619\backslash \nonumber\\
&&\hspace{22em}2274173103880744339809,\nonumber\\
&&C_{6+\phantom{1}}=
0.118322588863244285519212856456397718968975725227410541191067925,\nonumber\\
&&C_{10-}=0.464207706785944087396503330097938832697360392193891710489569762,\nonumber\\
&&C_{10+}=0.0123440983021588166317669811773152519959150566201343.
\end{eqnarray}
We have not yet been able to identify these constants but expect them to be of a similar status to the constants $C_{0\pm}$ in (\ref{5}) which are related to solutions of the Painlev\'e III \cite{Wu76,BMW73,TM73} or Painlev\'e V equation \cite{JM17,AP2}. We note that the constants must relate to the expansion coefficients in (2.27) of \cite{Kong}, which have to satisfy a Painlev\'e V hierarchy of differential equations and should lead to further coefficients $C_{12,\pm}$, $C_{14,\pm}$, etc. We also note that in (\ref{fnew}) we have split off a factor $k^{1/4}$, leaving only even powers of $\tau$ in the expansions of $F/k^{1/4}$.

The staggered susceptibility at the ferromagnetic point of a bipartite lattice, or what is equivalent, the susceptibility for an antiferromagnet, is given by an expression of the same form as (\ref{chiform}). For the square lattice the $F_{\pm}|^{\mathrm{af}}$ vanishes; there is only a background $B^{\mathrm{af}}$ as found in \cite{ONGP}. On the other hand the Fisher \cite{Fisher} relation 
\begin{equation}
\bar\chi_\pm^{\mathrm{hc}}|^{\mathrm{af}}=
2\bar\chi_\pm^{\mathrm{tr}}-\bar\chi_\pm^{\mathrm{hc}}
\end{equation}
together with (\ref{6}) and (\ref{fnew}) implies that if we define
\begin{equation}
C_{0\pm}^{\mathrm{hc}}|^{\mathrm{af}} = 8 C_{0\pm}/\sqrt{27}
\end{equation}
then
\begin{equation}\label{11}
F_\pm^{\mathrm{hc}}|^{\mathrm{af}} = F_\pm^{\mathrm{tr}} - F_\pm^{\mathrm{hc}} = k^{1/4} \left[ - C_{6\pm} \tau^6 + C_{6\pm} \tau^8 - (C_{6\pm} - C_{10\pm}) \tau^{10} + \mathrm{O}(\tau^{12}) \right].
\end{equation}
Also,
\begin{equation}\label{12}
B^{\mathrm{hc}}|^{\mathrm{af}} = 2 B^{\mathrm{tr}} - B^{\mathrm{hc}}.
\end{equation}
Equations (\ref{11}) and (\ref{12}) have provided significant tests confirming the correctness and accuracy of our numerical analyses. 

To the constants $C_{6\pm}$ in (\ref{8}) one could add the same rational above and below $T_{\mathrm{c}}$ and, on absorbing this change in other rationals in (\ref{fnew}), leave those equations unchanged in form. A corresponding replacement $C_{10\pm} \rightarrow C_{10\pm} + \mbox{ rational } \times C_{6\pm} + $ rational with similar consequences is possible. This non-uniqueness in (\ref{fnew}) has been removed by arbitrarily adopting the particularly simple form for $F_\pm^{\mathrm{hc}}|^{\mathrm{af}}$ in (\ref{11}). Note however that any such redefinitions can never eliminate the irrationals from (\ref{fnew}) or (\ref{11}) and we conclude that this is evidence for at least two irrelevant scaling fields beyond the one breaking rotational invariance and contributing first at O($\tau^4$) to the square lattice susceptibility. Furthermore, the presence of the same irrationals in the scaling functions in (\ref{fnew}) is evidence for a universality in terms beyond the leading order. We will elaborate on this in section \ref{sec:scaling} where, among other things, we make comparisons with the Aharony and Fisher \cite{AF83,AF80} scaling functions. 

It is also possible, based on existing results, to derive the reduced susceptibility of the Ising model on the kagom\'e lattice. This is given by \cite[eqn.~2.1]{Gut}, in terms of the reduced susceptibility of the model on the honeycomb lattice. Further aspects of this connection can be found in \cite{GG78}. With $Q = J/k_{\mathrm{B}}T$ for the kagom\'e lattice and $z = 2/(\mathrm{e}^{4Q}+1)$, this equation can be written as
\begin{equation}\label{kagome}
\bar\chi^{\mathrm{ka}} = \frac{3}{2} (1 - z^2)\bar\chi^{\mathrm{hc}} + \frac{1}{2} \left( (1+z^2) - (1-z^2) \langle \sigma_i \sigma_j \rangle_{\mathrm{nn}}^{\mathrm{hc}} \right).
\end{equation}

We note that the $z$ variable is the same variable as in (\ref{k}), pertaining to the interaction strength on the honeycomb lattice that results from reversing the star-triangle and decoration transformations on the kagom\'e lattice. We can also associate with the kagom\'e lattice the elliptic parameter and temperature variable associated with the honeycomb lattice, as given in (\ref{3}) and (\ref{k}). The average $\langle\sigma_i\sigma_j \rangle_{\mathrm{nn}}^{\mathrm{hc}}$ in (\ref{kagome}) is the nearest-neighbour correlation function of the honeycomb lattice, which is a simple multiple of the internal energy. It is given explicitly by eq.~(\ref{h1}) below.

As the second term in (\ref{kagome}) is a ``short-distance" term, it does not contribute to the scaling function $F_\pm^{\mathrm{ka}}$, which is thus entirely derived from the first term in (\ref{kagome}). By absorbing an extra normalising factor associated with $1-z^2$ into the constant term, we derive
\begin{eqnarray}\label{kagomesc}
C_{0\pm}^{\mathrm{ka}} & = & (-9+6 \sqrt{3}) C_{0\pm}^{\mathrm{hc}}, \nonumber \\
F_{\pm}^{\mathrm{ka}} & = & \frac{1-z^2}{1-z_{\mathrm{c}}^2} F_{\pm}^{\mathrm{hc}} \\
	& = & \left(1+\left(-1+\frac{\sqrt{3}}{2}\right)\tau + \left(1-\frac{5\sqrt{3}}{8}\right) \tau^2+\left(-\frac{11}{16}+\frac{13\sqrt{3}}{32}\right) \tau^3 + \ldots\right) F_{\pm}^{\mathrm{hc}}, \nonumber \\
B^{\mathrm{ka}} & = & \frac{3}{2} (1 - z^2) B^{\mathrm{hc}} + \frac{1}{2} \left( (1+z^2) - (1-z^2) \langle \sigma_i \sigma_j \rangle_{\mathrm{nn}}^{\mathrm{hc}} \right), \nonumber
\end{eqnarray}
where $F_{\pm}^{\mathrm{hc}}$ is given in (\ref{fnew}) and $z_{\mathrm{c}} = 2 - \sqrt{3}$.
The two leading terms of $\chi^{\mathrm{ka}}$ near $T_{\mathrm{c}}$ were studied before in connection with generalised extended lattice-lattice scaling \cite{Gut,GG78,AP03}.

\section{Scaling theory and CFT predictions\label{sec:scaling}}

\subsection{Scaling theory\label{sec:scalingtheory}}
The singular part of the dimensionless free energy\footnote{In the following, we shall use the notation $f=\log z=-\beta\Psi$, with $z$ the partition function per site and $\Psi$ the usual free energy per site.} of the two-dimensional Ising model satisfies the following scaling Ansatz:
\begin{eqnarray}\label{scaling}
f_{\mathrm{sing}}(g_t,g_h,\{g_{u_j}\}) &=& -g_t^2\log{|g_t|}\cdot {\tilde Y}_{\pm}(g_h/|g_t|^{y_h/y_t},\{g_{u_j}/|g_t|^{y_j/y_t}\}) \\ \nonumber
&&+\; g_t^2\cdot { Y}_{\pm}(g_h/|g_t|^{y_h/y_t},\{g_{u_j}/|g_t|^{y_j/y_t}\}).
\label{s1}
\end{eqnarray}
Here $g_t, \,\, g_h, \,\, g_{u_j}$ are nonlinear scaling fields associated, respectively, with the thermal field $\tau$, the magnetic field $h$ and the irrelevant fields ${u_j}$.\footnote{The scaling function $Y_{\pm}(x,\{0\})$, without the effects of irrelevant fields, has been studied recently to high precision, see \cite{MDMB} and references cited therein.} The exponent $y_t $ is the thermal exponent, and takes the value $1$ for the two-dimensional Ising model, while $y_h$ is the magnetic exponent and takes the value $15/8.$ The irrelevant exponents $y_j$ are all negative. In the language of conformal field theory (CFT), this scaling Ansatz assumes only a single resonance between the identity and the energy. That the dimensions are integers implies that there might be multiple resonances which give rise higher powers of $\log \tau$ as observed. We have not included such terms in the scaling Ansatz above, as there are no $g_t^2(\log|g_t|)^n$ terms with $n>1$. Following our earlier analysis \cite{ONGP}, Caselle et al.\ \cite{C02} discussed the scaling theory of the two-dimensional Ising model in considerable depth, in particular the conclusions that could be drawn about the irrelevant operators. We discuss this further below.

The nonlinear scaling fields have power series expansions with coefficients which are smooth functions of $\tau$ and the irrelevant variables $u \equiv \{u_j\}.$ In particular one has
\begin{eqnarray}
\label{s2}
g_t &=& \sum_{n \ge 0} a_{2n}(\tau,u) \cdot h^{2n}, \, \,\,\, a_0(0,u) = 0, \\ \nonumber
g_h &=& \sum_{n \ge 0} b_{2n+1}(\tau,u) \cdot h^{2n+1},\\ \nonumber
g_{u_j} &=& \sum_{n \ge 0} c_{2n}(\tau,u) \cdot h^{2n}. \nonumber
\end{eqnarray}

In the absence of irrelevant fields, the known zero-field free energy imposes the equalities ${\tilde Y}_{+}(0)={\tilde Y}_{-}(0)$ and $Y_{+}(0)=Y_{-}(0)$. Furthermore, the known solution for the magnetisation, which contains no logarithms, and the known (but not proved) absence of logarithmic terms in the divergent part of the susceptibility impose the constraints that the first and second derivatives of ${\tilde Y}_{\pm}(0)$ also vanish. That is to say, ${\tilde Y}^{'}_{\pm}(0)={\tilde Y}^{''}_{\pm}(0)=0$. Aharony and Fisher \cite{AF83} have argued, almost certainly correctly, that there are no logarithms multiplying the leading power law divergence of all higher order field derivatives, not just the first two, as discussed. In that case it follows that ${\tilde Y}_{\pm}$ are constants, and further the analyticity on the critical isotherm for $h \ne 0$ requires high-low temperature equality, ${\tilde Y}_{+}={\tilde Y}_{-}$. Collecting all this information, we have, for the zeroth, first and second field derivatives of the free energy,
\begin{eqnarray}
\label{s3}\, 
f(\tau,h=0) &=& -A \,(a_0(\tau))^2\, \log{|a_0(\tau)|} + A_0(\tau), \nonumber \\ 
{\mathcal M}(\tau < 0, h=0) &=& B\,b_1(\tau)\, |a_0(\tau)|^{\beta}, \\  \nonumber
k_{\mathrm{B}}T\chi_{\pm}(\tau, h=0) &=& C_{{\pm}}(b_1(\tau))^2 \, |a_0(\tau)|^{-\gamma} - E \,a_2(\tau)\, a_0(\tau) \, \log{|a_0(\tau)|} + D(\tau),
\end{eqnarray}
where $A$, $B$, $C_{\pm}$ and $E$ are constants, the background term $A_0(\tau)$ is a power series in $\tau$, and the critical exponents are $\beta=1/8$ and $\gamma=7/4$. The free energy and magnetisation determine the scaling field coefficients $a_0(\tau)$ and $b_1(\tau)$ which, given our freedom in choice of $A$ and $B$, can be normalized to $a_0(\tau)=\tau+\mathrm{O}(\tau^2)$ and $b_1(\tau)=1 +\mathrm{O}(\tau)$. The presence of any irrelevant scaling fields will manifest themselves as deviations in the predicted form of the susceptibility in (\ref{s3}).\footnote{According to (\ref{background}), the background contribution $D(\tau)$ contains terms with arbitrary powers of $\log|\tau|$, which have not yet been interpreted within the context of scaling theory.}

To get an explicit expression for the predicted susceptibility in the absence of irrelevant fields we start with the zero field magnetization which is known to be the same function 
\begin{equation}\label{16}
M = (1 - k^2)^{1/8} = 2^{1/4} k^{1/8} (1+\tau^2)^{1/16} (-\tau)^{1/8}
\end{equation}
for all three (square, triangular and honeycomb) lattices. The second equality in (\ref{16}) follows from our temperature definition (\ref{3}) and if we use this to solve for $b_1(\tau)$ in (\ref{s3}) we can reduce the zero field susceptibility in (\ref{s3}) to 
\begin{equation}\label{17}
k_{\mathrm{B}} T \chi_\pm = C_\pm |\tau|^{-7/4} F_\pm - E a_2(\tau) a_0(\tau) \log |a_0(\tau)| + D(\tau),
\end{equation}
where
\begin{equation}\label{18}
F_\pm = k^{1/4} (1+\tau^2)^{1/8} (\tau/a_0(\tau))^2.
\end{equation}
It only remains to determine $a_0(\tau)$ from the singular part of the zero field free energy for each lattice to complete the calculation of $F_\pm$ which we henceforth denote as the Aharony and Fisher scaling function $F_\pm(\mathrm{A\&F})$. 

It will turn out to be useful\footnote{Identities used can be found in \cite{GR}, see eqs.\ 2.597.1, 8.112.3, 8.113.1, 8.113.3, 8.126.3 and 9.131.1.} to define the following integral, in terms of which the internal energy is defined:
\begin{equation}
\label{I}
I(\tau) = \frac{2}{\pi}  \int_0^{\pi/2} \frac{d\theta}{\sqrt{\tau^2 + \sin^2 \theta}} = \frac{2}{\pi \sqrt{1+\tau^2} }\mathrm{K} \Big( \frac{1}{\sqrt{1+\tau^2}} \Big) =  \frac{4\sqrt{k}}{\pi (1+k) }\mathrm{K} \Big( \frac{2\sqrt{k}}{1+k} \Big),
\end{equation}
where $\mathrm{K}$ is the complete elliptic integral of the first kind. This function is invariant under the high-low temperature change $k \to 1/k$. Useful forms at both high and low temperatures are obtainable from the Landen transformation,
\begin{equation}\label{Landen}
\mathrm{K} \Big( \frac{2\sqrt{k}}{1+k} \Big) = (1+k) \mathrm{K}(k) = \Big(1 + \frac{1}{k}\Big) \mathrm{K}\Big(\frac{1}{k}\Big).
\end{equation}
For the subsequent scaling analysis we will require the singular part of $I(\tau),$ which is 
\begin{eqnarray}
I(\tau)_{\mathrm{sing}} &=& -\frac{2}{\pi \sqrt{1+\tau^2} }\log {|\tau|}\cdot \mathrm{K} \Big( \frac{\tau}{\sqrt{1+\tau^2}} \Big) \\ \nonumber
&=& -\frac{\log {|\tau|}}{ \sqrt{1+\tau^2} }\cdot {_2F_1}\left (\frac{1}{2},\frac{1}{2};1;\frac{\tau^2}{1+\tau^2}\right ) = -\log {|\tau|}\,\cdot {_2F_1}\left (\frac{1}{2},\frac{1}{2};1;-\tau^2 \right ).
\end{eqnarray}
We next write the internal energy, per site, in terms of the above integral (\ref{I}):
For the square lattice, 
\begin{equation}
\label{sq1}
\frac{\partial f}{\partial K}\bigg|_{\mathrm{sq}} = 2 \langle \sigma_i \sigma_j \rangle_{\mathrm{nn}} = \coth({2K_{\mathrm{sq}}})(1 - \tau I(\tau)),\end{equation}
where $f=-\beta\Psi$ with $\Psi$ the free energy per site.
For the triangular lattice, 
\begin{equation}
\label{tr1}
\frac{\partial f}{\partial K}\bigg|_{\mathrm{tr}} = 3 \langle \sigma_i \sigma_j \rangle_{\mathrm{nn}} =\frac{1+u}{1-u}\left (1-\frac{3u-1}{2[u^3(1-u)^3(1+3u)]^{1/4}} I(\tau)\right ).\end{equation}
For the honeycomb lattice, 
\begin{equation}
\label{h1}
\frac{\partial{f}}{\partial K}\bigg|_{\mathrm{hc}} = \frac{3}{2} \langle \sigma_i \sigma_j \rangle_{\mathrm{nn}} =\frac{1+z^2}{1-z^2}\left (1-\left (\frac{1+z}{1-z}\right )^{3/2} \frac{4z-1-z^2}{8[z^3(1-z+z^2)]^{1/4}} I(\tau)\right ).\end{equation}
We can calculate the zero-field free energy by integrating these expressions. In fact we are only interested in the singular part of the free energy, which we normalise by the requirement that it vanishes at $T_{\mathrm{c}}.$ With that normalisation, we can write
\begin{equation}\label{24}
f_{\mathrm{sing}} = - \log |\tau| \int_0^\tau d\tau \frac{dK}{d\tau} C_I \cdot {_2F_1}\left(\frac{1}{2}, \frac{1}{2};1;-\tau^2\right)
\end{equation}
which is to be compared to $f_{\mathrm{sing}} = -A a_0(\tau)^2 \log|\tau|$ in (\ref{s3}). The $C_I$ in (\ref{24}) is the coefficient of $I(\tau)$ in equations (\ref{sq1})--(\ref{h1}) for the appropriate lattice; the $dK/d\tau$ is also to be evaluated with $K$ for the appropriate lattice.

For the square lattice we determine $2(dK_{\mathrm{sq}}/d\tau)C_I = \tau/\sqrt{1+\tau^2}$ so that the integrand in (\ref{24}) is seen to be explicitly odd in $\tau$ and we can identify $A^{\mathrm{sq}}=1/4$ and the even function 
\begin{equation}\label{25}
a_0(\tau)^2\big|_{\mathrm{sq}} = \tau^2 \left( 1 - \frac{3}{8} \tau^2 + \frac{41}{192} \tau^4 - \frac{147}{1024} \tau^6 + \frac{8649}{81920} \tau^8 - \frac{10769}{131072} \tau^{10} + \mathrm{O}(\tau^{12}) \right).
\end{equation}
Equation (\ref{25}) combined with (\ref{18}) gives
\begin{equation}
F_\pm(\mathrm{A\&F})^{\mathrm{sq}} = k^{1/4} \left[ 1 + \frac{1}{2} \tau^2 - \frac{31}{384} \tau^4 + \frac{125}{3072} \tau^6 - \frac{38147}{1474560} \tau^8 + \frac{108713}{5898240} \tau^{10} + \mathrm{O}(\tau^{12}) \right]
\end{equation}
which extends the result in \cite{ONGP} to higher order.

For the triangular lattice we find $8(dK_{\mathrm{tr}}/d\tau)C_I/\sqrt{27} = \tau -\tau^3/2+97\tau^5/256+\ldots = \sum c_n \tau^{2n+1}$ with the $c_n$ satisfying the three term recursion $(n^2+n+2/9)c_n +(2n^2-11/18)c_{n-1} +(n^2-n-11/18-15/(144(n^2-n)))c_{n-2}=0$. The integrand in (\ref{24}) is again odd in $\tau$ and we can identify $A^{\mathrm{tr}}=\sqrt{27}/16$. For the honeycomb lattice $A^{\mathrm{hc}}=\sqrt{27}/32$; otherwise the integrand is the same. The scaling functions are \begin{equation}
a_0(\tau)^2 |_{\mathrm{tr,hc}} = \tau^2 \left(1 - \frac{3}{8} \tau^2 + \frac{55}{256} \tau^4 - \frac{149}{1024} \tau^6 + \frac{17667}{163840} \tau^8 - \frac{44321}{524288} \tau^{10} + \mathrm{O}(\tau^{12}) \right)
\end{equation}
and
\begin{equation}
F_\pm(\mathrm{A\&F})^{\mathrm{tr,hc}} = k^{1/4} \left[ 1 + \frac{1}{2} \tau^2 - \frac{21}{256} \tau^4 + \frac{85}{2048} \tau^6 - \frac{8669}{327680} \tau^8 + \frac{49507}{2621440} \tau^{10} + \mathrm{O}(\tau^{12}) \right].
\end{equation}

We can now compare these scaling functions based on the assumption of no corrections to scaling with the observed functions given in (\ref{fnew}). Define $\Delta F_\pm = F_\pm - F_\pm(\mathrm{A\&F})$; then
\begin{eqnarray}\label{29}
\Delta F_\pm^{\mathrm{sq}} & = & k^{1/4} \left[ -\frac{\tau^4}{384} + \left(\frac{11}{7680} - \frac{7 C_{6\pm}}{5} \right) \tau^6 - \left( \frac{21421}{22118400} - \frac{4973 C_{6\pm}}{3600} \right) \tau^8 \right. \nonumber \\
	& & \hspace{1cm} \left. + \left( \frac{894191}{1238630400} - \frac{100261 C_{6\pm}}{115200} - \frac{793 C_{10\pm}}{210} \right) \tau^{10} + \mathrm{O}(\tau^{12}) \right], \nonumber \\
\Delta F_\pm^{\mathrm{tr}} & = & k^{1/4} \left[ - \frac{3 C_{6\pm}}{2} \tau^6 - \left( \frac{1}{102400} - \frac{1209 C_{6\pm}}{800} \right) \tau^8 \right. \nonumber \\
	& & \hspace{1cm} \left. + \left( \frac{43}{2867200} - \frac{261 C_{6\pm}}{200} - \frac{51 C_{10\pm}}{70} \right) \tau^{10} + \mathrm{O}(\tau^{12}) \right], \\
\Delta F_\pm^{\mathrm{hc}} & = & k^{1/4} \left[ - \frac{C_{6\pm}}{2} \tau^6 - \left( \frac{1}{102400} - \frac{409 C_{6\pm}}{800} \right) \tau^8 \right. \nonumber \\
	& & \hspace{1cm} \left. + \left( \frac{43}{2867200} - \frac{61 C_{6\pm}}{200} - \frac{121 C_{10\pm}}{70} \right) \tau^{10} + \mathrm{O}(\tau^{12}) \right]. \nonumber
\end{eqnarray}
The absence of a correction at O($\tau^4$) in $F_\pm^{\mathrm{tr}}$ and $F_\pm^{\mathrm{hc}}$ is expected since the operator that breaks rotational invariance on the square lattice is not present on these lattices. On the other hand Caselle et al.\ also suggested that because the operator that breaks rotational invariance on the triangular lattice first contributes at O($\tau^8$) there might not be any O($\tau^6$) correction. The clear evidence in (\ref{29}) of such a correction on the triangular lattice, and indeed on all three lattices, shows that there are corrections to scaling operators in the Ising model that are not associated just with the breaking of rotational invariance. We elaborate on this in the following section. 

In a similar manner, we can derive the Aharony and Fisher scaling function for the kagom\'e lattice. The extra $1-z^2$ term in $\chi^{\mathrm{ka}}$ arises because the magnetization for the kagom\'e lattice is given by (\cite[eqn.~95]{Syo})
\begin{equation}
M = \sqrt{1-z^2} \left(1 - k^2\right)^{1/8},
\end{equation}
which replaces (\ref{16}). This introduces an extra $\sqrt{1-z^2}$ factor into $b_1(\tau)$ and (\ref{18}) becomes
\begin{equation}
F_\pm = \frac{1-z^2}{1-z_{\mathrm{c}}^2} k^{1/4} (1+\tau^2)^{1/8} (\tau/a_0(\tau))^2,
\end{equation}
where the denominator in the first term is a normalising factor with
$z_{\mathrm{c}}=2-\sqrt{3}$ the critical value on the honeycomb lattice.
The remainder of the derivation of the A \& F scaling function is unchanged, resulting in the relation
\begin{equation}\label{afkagome}
F_\pm(\mathrm{A\&F})^{\mathrm{ka}} = \frac{1-z^2}{1-z_{\mathrm{c}}^2} F_\pm(\mathrm{A\&F})^{\mathrm{hc}}.
\end{equation}
In view of (\ref{kagomesc}) and (\ref{afkagome}), we also know that the deviation of the kagom\'e lattice scaling function from the corresponding A \& F scaling function is identical to that of the honeycomb up to a factor,
\begin{equation}
\Delta F_\pm^{\mathrm{ka}} = \frac{1-z^2}{1-z_{\mathrm{c}}^2} \Delta F_\pm^{\mathrm{hc}}.
\end{equation}

\subsection{Scaling from conformal field theory\label{sec:conformal}}

This section draws extensively on the paper by Caselle et al.\ \cite{C02} which was written after the appearance of \cite{ONGP}. We adopt the usual notation within CFT. At the critical point, the Ising model is describable by the unitary minimal CFT with central charge $c = 1/2.$ The spectrum can be divided into three conformal families. They are the identity, spin and energy families, commonly denoted $[I],$ $[\sigma],$ and $[\epsilon]$ respectively. Each family characterizes a different transformation property under the dual and ${\mathbb Z}_2$ symmetries.  $T$ denotes the energy-momentum tensor, so $T\bar{T}$ is a spin-zero irrelevant operator. Each family contains one {\it primary field} and a number of {\it secondary fields.} The conformal weights of the primary fields are $h_I = 0,$ $h_{\sigma} = 1/16,$ and $h_{\epsilon} = 1/2,$ and all primary fields are relevant.

The secondary fields are derived from the primary fields by applying the generators $L_{-i}$ and $\bar{L}_{-i}$ of an appropriate Virasoro algebra. $L_{-1}$ plays a particular role, being the generator of translations on the lattice, and so gives zero acting on any translationally invariant observable. Another important concept is that of a {\it quasi-primary operator.} A quasi-primary field $|Q \rangle$ is a secondary field satisfying $L_1|Q \rangle = 0.$ This condition eliminates all secondary fields generated by $L_{-1}.$ As quasi-primary operators are the only ones which can appear in translationally invariant quantities, they played a central role in the analysis of Caselle et al.\ \cite{C02}, and also in our current analysis, as they are the natural candidates for irrelevant operators.

To make the connection between the scaling Ansatz given in eqn. (\ref{scaling}) and the discussion in terms of CFT, we first, for simplicity, set $y_t$ to its numerical value, 1, and replace the scaling field $g_t$ by its leading term $\tau.$ Then the terms $Y_{\pm}$ and ${\tilde Y}_{\pm}$ in eqn. (\ref{scaling}) can be easily expanded. They will involve terms of the form
\begin{equation}
\prod_i \left ( \frac{g_i}{|\tau|^{y_i}} \right )^{p_i}=\prod_{i \in \sigma} \left ( \frac{g_{\sigma_i}}{|\tau|^{y_{\sigma_i}}} \right )^{p_i}\cdot \prod_{i \in I} \left ( \frac{g_{I_i}}{|\tau|^{y_{I_i}}} \right )^{p_i} \cdot \prod_{i \in \epsilon} \left ( \frac{g_{\epsilon_i}}{|\tau|^{y_{\epsilon_i}}} \right )^{p_i}.
\end{equation}
As the susceptibility is the second field derivative of the free energy, we must retain terms with exactly two factors in the first of the three products above, that is, terms of the form
\begin{equation}
\label{susterms}
 \frac{g_{\sigma_1}\cdot g_{\sigma_2}}{|\tau|^{y_{\sigma_1} }\cdot |\tau|^{y_{\sigma_2}}}  \cdot \prod_{i \in I} \left ( \frac{g_{I_i}}{|\tau|^{y_{I_i}}} \right )^{p_i} \cdot \prod_{i \in \epsilon} \left (\frac{g_{\epsilon_i}}{|\tau|^{y_{\epsilon_i}}} \right )^{p_i}.
\end{equation}

Recall the prefactor $g_t^2 \sim \tau^2$ before the terms $Y_{\pm}$ and ${\tilde Y}_{\pm}$ in eqn. (\ref{scaling}). Including this prefactor, it is clear that all terms of order $\tau^N$ in the susceptibility are given by all terms in eqn (\ref{susterms}) satisfying
\begin{equation}
N = 2 - (y_{\sigma_1} + y_{\sigma_2} + \sum p_i y_{I_i} + \sum p_i y_{\epsilon_i}).
\label{Nmin2}
\end{equation}
The leading term in the susceptibility occurs when there are no $\epsilon$ or $I$ fields and $y_{\sigma_1} = y_{\sigma_2} = y_h = 15/8,$ giving $N=2-15/8-15/8 = -7/4,$ which is the well-known susceptibility exponent. Exponents for other terms in the table rely on eigenvalue exponents given by Caselle et al.\ \cite{C02}, which we summarise in Table \ref{table1}.

\begin{table}
\begin{tabular}{|r|l|l|} \hline \hline

Eigenvalue & Term & Term  \\ \hline
 & &  \\
$-2$ & $Q_2^{I} \bar{Q}_2^{I}=T\bar {T}$ & $Q_4^{I} + \bar{Q}_4^{I} $  (sq)  \\
$-3$ & $Q_4^{\epsilon} + \bar{Q}_4^{\epsilon}$ (sq) &  \\
$-4$ & $Q_6^{I} + \bar{Q}_6^{I}$ (tr) &  \\
$-5$ & $Q_6^{\epsilon} + \bar{Q}_6^{\epsilon}$ (tr) &  \\
$-6$ & $Q_4^I \bar{Q}_4^I$ & $Q_8^{I} + \bar{Q}_8^{I}$ (sq)   \\
$-7$ & $Q_4^{\epsilon}  \bar{Q}_4^{\epsilon} $ & $Q_8^{\epsilon} + \bar{Q}_8^{\epsilon} $ (sq)   \\
$-8$ &  &   \\
$-10$ & $Q_{12}^{I} + \bar{Q}_{12}^{I}$  &$Q_6^{I} \bar{Q}_6^{I}$  \\
$-4\frac{1}{8}$ & $Q_3^{\sigma} \bar{Q}_3^{\sigma}$ & $Q_6^{\sigma} + \bar{Q}_6^{\sigma}$ (tr)  \\
$-6\frac{1}{8}$ & $Q_8^{\sigma} + \bar{Q}_8^{\sigma}$ (sq)&  \\
$-8\frac{1}{8}$ & $Q_5^{\sigma} \bar{Q}_5^{\sigma}$ & $Q_3^{\sigma}\bar{Q}_7^{\sigma} + {Q}_7^{\sigma}\bar{Q}_3^{\sigma}$ (sq)  \\
 & &  \\
\hline \hline
\end{tabular}
\caption{Eigenvalues of various operator combinations that contribute to the susceptibility. The spin-zero and spin-12 operators (unlabelled) contribute to both square and triangular lattices. The spin-4 and spin-8 operators contribute only to the square lattice (labelled (sq)), while the spin-6 operators, labelled (tr), contribute only to the triangular lattice.}
\label{table1}
\end{table}

Caselle et al.\ \cite{C02} have produced a list of irrelevant operators and we reproduce combinations of these operators that contribute to $\chi^{\mathrm{sq}}$ and $\chi^{\mathrm{tr}}$ together with the primary spin operator $\sigma$ in Table \ref{operator}. Power counting as described in \cite{C02} and above and leading to relation (\ref{Nmin2}), determines when each combination first contributes. Because corrections at O($\tau^2$) in both $F_\pm^{\mathrm{sq}}$ and $F_\pm^{\mathrm{tr}}$ are not observed and similarly corrections at O($\tau^4$) in $F_\pm^{\mathrm{tr}}$ are absent, we adopt the assumption of Caselle et al.\ that all contributions from combinations of the form\footnote{Here and elsewhere we adopt the convention that $O^x$ is a generic operator in family $[x]$.} $\sigma^2 O^I O^\varepsilon (T \bar{T})^n$, $n > 0$, vanish, as well as all descendants of $\sigma^2(T{\bar T})^n,$ and consequently these entries are excluded from Table \ref{operator}.

\begin{table}\renewcommand{\arraystretch}{1.2}
\begin{tabular}{|l|l|l|} \hline \hline
N & Square & Triangular \\ \hline
& & \\
0 & $\sigma^2$ & $\sigma^2$ \\   
& & \\
2 & $---$ & $---$\\  
& & \\ 
4 & $\sigma^2(Q_4^{I} + \bar{Q}_4^I)^2$ & $---$ \\ 
& & \\
6 & $\sigma^2(Q_4^{\epsilon} + \bar{Q}_4^{\epsilon})^2$ & $\sigma^2(Q_4^I \bar{Q}_4^I)$ \\ 
 & $\sigma^2(Q_4^I \bar{Q}_4^I)$ & $\sigma(Q_3^{\sigma} \bar{Q}_3^{\sigma})$ \\
 &  $\sigma(Q_3^{\sigma} \bar{Q}_3^{\sigma})$ &  \\ 
 & & \\
8 & $\sigma^2(Q_4^{I} + \bar{Q}_4^{I})^4$ & $\sigma^2(Q_6^{I} + \bar{Q}_6^{I})^2$  \\ 
  & & \\
  10 & $\sigma^2(Q_4^{I} + \bar{Q}_4^{I})^2(Q_4^{\epsilon} + \bar{Q}_4^{\epsilon})^2$ & $\sigma^2(Q_6^{\epsilon} + \bar{Q}_6^{\epsilon})^2$  \\ 
    & $\sigma^2(Q_4^{I} + \bar{Q}_4^{I})^2(Q_4^{I} \bar{Q}_4^{I})$ & $\sigma^2(Q_6^{I} \bar{Q}_6^{I})$  \\ 
    & $\sigma^2(Q_6^{I} \bar{Q}_6^{I})$ & $\sigma(Q_5^{\sigma} \bar{Q}_5^{\sigma})$   \\
   & $\sigma(Q_4^{I} + \bar{Q}_4^{I})^2(Q_3^{\sigma} \bar{Q}_3^{\sigma})$ &  \\   
    &  $\sigma(Q_5^{\sigma} \bar{Q}_5^{\sigma})$ & \\
  & & \\  
   12 & $\sigma^2O^IO^{\epsilon}$ (many terms) & $\sigma^2O^IO^{\epsilon}$ (many terms) \\
    & $\sigma(Q_4^{\epsilon} + \bar{Q}_4^{\epsilon})^2(Q_3^{\sigma} \bar{Q}_3^{\sigma})$ & $\sigma(Q_3^{\sigma} \bar{Q}_3^{\sigma})(Q_4^{I} \bar{Q}_4^{I})$  \\   
 &$\sigma(Q_3^{\sigma} \bar{Q}_3^{\sigma}) (Q_4^{I} \bar{Q}_4^{I}) $ & $\sigma(Q_6^{\sigma} \bar{Q}_6^{\sigma}) $\\
 &  $\sigma(Q_6^{\sigma} \bar{Q}_6^{\sigma})$ & $(Q_6^{\sigma} + \bar{Q}_6^{\sigma}) ^2$ \\
  & $(Q_3^{\sigma} \bar{Q}_3^{\sigma})^2$  & $(Q_3^{\sigma} \bar{Q}_3^{\sigma})^2 $ \\
  & & \\
      \hline \hline
\end{tabular}
\caption{Operator combinations contributing to the susceptibility. The $N$ values in the first column specify the leading contribution $|\tau|^{-7/4+N}$ to $\chi^{\mathrm{sq}}$ and $\chi^{\mathrm{tr}}$ or $|\tau|^N$ to $F_\pm^{\mathrm{sq}}$ and $F_\pm^{\mathrm{tr}}$ of the corresponding entries in the second and third columns.}\label{operator}
\end{table}

Because there are multiple operator combinations at most correction levels in Table \ref{operator}, a unique identification of correction terms with operators is in general not possible. Thus the following remarks are to be viewed either as pure speculation or at best a set of assumptions consistent with the corrections to scaling that are displayed in (\ref{29}).

\begin{enumerate}
\item The corrections observed in (\ref{29}) are consistent with the conjecture that all operator combinations of the form $O^\sigma O^I$, $O^\sigma O^\varepsilon$ or $O^\sigma O^I O^\varepsilon$ are rational multiples of the leading order contribution of $O^\sigma$. Furthermore these multipliers are the same above and below $T_{\mathrm{c}}$. This makes these contributions particularly hard to distinguish from the scaling fields associated with the leading contribution. For example, the rational coefficient 11/7680 of $\tau^6$ in $\Delta F_\pm^{\mathrm{sq}}$ in (\ref{29}) is very likely a combination of a direct contribution from $\sigma^2 (Q_4^\varepsilon + \bar{Q}_4^\varepsilon)^2$ and a scaling field correction from the $\sigma^2 (Q_4^I + \bar{Q}_4^I)^2$ term, whose leading contribution is at order $\tau^4$. \label{comment1}

\item We identify all irrational corrections with $\sigma$-field operators. Specifically, contributions proportional to $C_{6\pm}$ with $\sigma(Q_3^\sigma \bar{Q}_3^\sigma)$ and those proportional to $C_{10\pm}$ with $\sigma(Q_5^\sigma \bar{Q}_5^\sigma)$. The ambiguity in $C_{6\pm}$ and $C_{10\pm}$ as discussed following eqns.\ (\ref{fnew})--(\ref{11}) is relevant in the present context. A part of $C_{6\pm}$ might be a rational number associated with $\sigma^2(Q_4^I \bar{Q_4}^I)$ and this would further complicate the interpretation of the 11/7680 coefficient of $\tau^6$ described in item \ref{comment1}. \label{comment2}

\item The coefficients of $C_{6\pm}$ in (\ref{29}) on the different lattices are, after dividing out the leading $\tau^6$ term, $1-4973\tau^2/5040+\ldots$ (square), $1-403\tau^2/400+\ldots$ (triangular) and $1-409\tau^2/400+\ldots$ (honeycomb). Because these are all different, we must conclude that the scaling function associated with $\sigma (Q_3^{\sigma} \bar{Q_3}^{\sigma})$ is lattice dependent. An analogy is the difference seen in the $F_{\pm}$ scaling function on the kagom\'e lattice as seen in (\ref{afkagome}). It is the equality of $F_{\pm}$ on the square, triangular and honeycomb lattices to $\mathrm{O}(\tau^3)$ that is to be considered as ``accidental" and not generic.

\item The very particular structure of the short-distance terms, given in (\ref{background}) is not explicitly predicted by CFT. Rather, since the primary logarithm, responsible for the specific heat behaviour, is due to a resonance between the thermal and identity operator \cite{C02}, we might expect additional multiple resonances, giving rise to higher powers of logarithms. These are indeed observed, but it does not appear to be possible to associate particular operators with these terms---at least not by our
naive method of just power counting.

\item Table \ref{operator} shows two new distinct $\sigma$-field operators at order $\tau^{12}$. If, as we have conjectured in item \ref{comment2}, each is associated with a new irrational $C_\pm$ then we can no longer make any unique identifications as we did for $C_{6\pm}$ and $C_{10\pm}$. For all terms in $F_\pm$ beyond $\tau^{10}$ we are left only with the numerical coefficients tabulated in appendix A.
\end{enumerate}

\section{Generation of series\label{sec:series}}

\subsection{Quadratic recurrences and {\mathversion{bold}${Z}$}-invariance}\label{jacques}

The algorithm deriving the susceptibility series for the isotropic
square lattice Ising model \cite{ONGP}, with $k=\sinh^2(2\beta J)$,
was rather simple using \cite{P}\footnote{For the uniform rectangular Ising model, using the methods of their
lattice-Painlev\'e III paper \cite{MWa}, McCoy and Wu \cite {MWb}
have generalized (\ref{Toda}) to the so-called $\lambda$-extended version,
in which the coefficient of $\lambda^j$ in $C(M,N;\lambda)$ is the $j$-particle contribution to the
pair correlation function $C(M,N)$. Equations like (\ref{Toda})
also exist for $n$-point correlation functions \cite{MPW}.}
\begin{eqnarray}
&&k[C(M,N)^2-C(M,N-1)C(M,N+1)]\nonumber\\
&&\qquad+[C^{\ast}(M,N)^2-C^{\ast}(M-1,N)C^{\ast}(M+1,N)]=0,\cr\nonumber\\
&&k[C(M,N)^2-C(M-1,N)C(M+1,N)]\nonumber\\
&&\qquad+[C^{\ast}(M,N)^2-C^{\ast}(M,N-1)C^{\ast}(M,N+1)]=0,
\label{Toda}
\end{eqnarray}
where
\begin{equation}
C(M,N)\equiv\langle\sigma_{0,0}\sigma_{M,N}\rangle,\quad
C^{\ast}(M,N)\equiv\langle\sigma_{0,0}\sigma_{M,N}\rangle^{\ast}
\end{equation}
with the asterisk denoting the corresponding quantities on the dual
lattice with the dual temperature obtained by replacing $k\to k^{\ast}=1/k$.
Series for the pair correlations can be solved iteratively using the
series for $C(1,0)=C(0,1)$ and the diagonal correlations $C(N,N)$ and
$C^{\ast}(N,N)$, which in turn follow from the well-known $C(0,0)=1$
and $C(1,1)$ by the Painlev\'e VI type iteration scheme of Jimbo and Miwa
\cite{JM17}, or with a little more work from the well-known Toeplitz determinants
\cite{MWbook}.

\begin{figure}[tbph]
\centerline{\includegraphics[width=200pt]{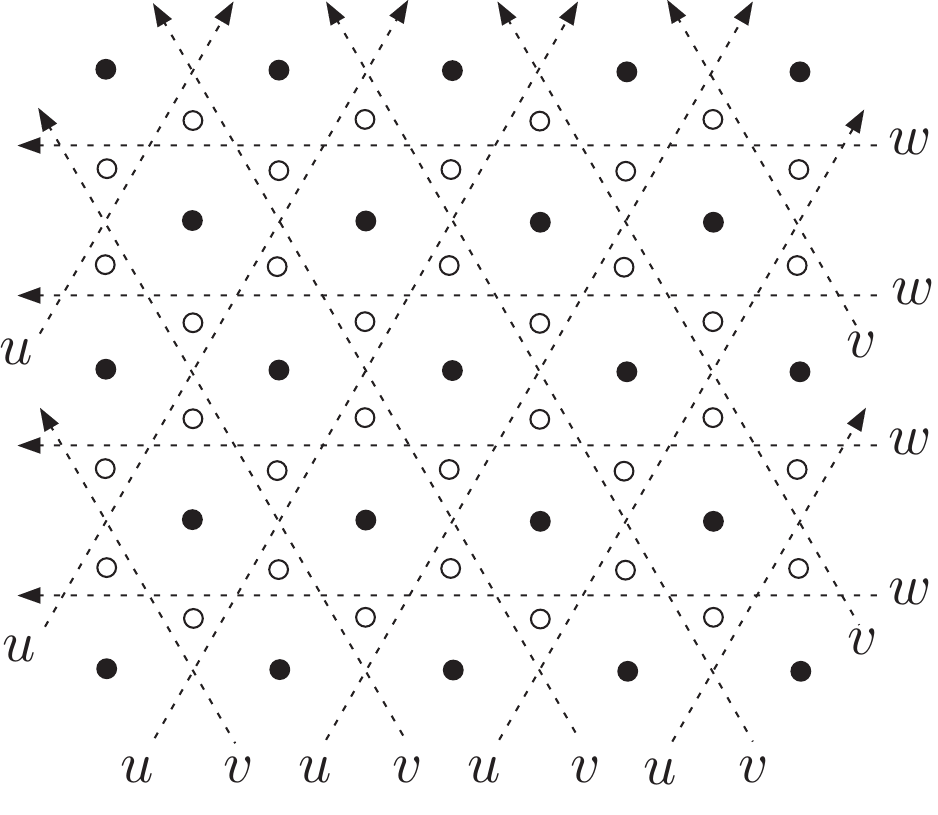}}
\caption{Parts of the infinite triangular lattice (black circles), honeycomb
lattice (open circles) and kagom\'e lattice of rapidity lines (oriented
dashed lines) with rapidities $u$, $v$ and $w$.}
\label{fig1}
\end{figure}

For the isotropic triangular and honeycomb lattices the situation is far more complicated. We have used the generalization of (\ref{Toda}) for
general planar lattices \cite{P}, together with Baxter's $Z$-invariance
\cite{B,APZI} as was first numerically implemented in \cite{AP1}.
More specifically, consider the situation in Figure \ref{fig1}:
The Ising model on the triangular lattice (black circles in the figure)
and its dual on the honeycomb lattice (open circles) are $Z$-invariant
in the sense of Baxter \cite{B}, with rapidity lines forming a kagom\'e
lattice (oriented dashed lines).\footnote{A kagom\'e Ising model can be obtained from the honeycomb Ising model by decoration and star-triangle
transformation \cite{Naya,Fisher} and its spins then live on all the
intersections of pairs of rapidity lines.}

To get the isotropic cases we need to choose the three rapidity
values as
\begin{equation}
u=\frac23{\rm K}(k'),\quad v=\frac13{\rm K}(k'),\quad w=0,
\label{rapidities}
\end{equation}
with $k'=\sqrt{1-k^2}$ and ${\rm K}(k)$ the complete elliptic integral
of the first kind.
\begin{figure}[tbph]
\centerline{\includegraphics[width=200pt]{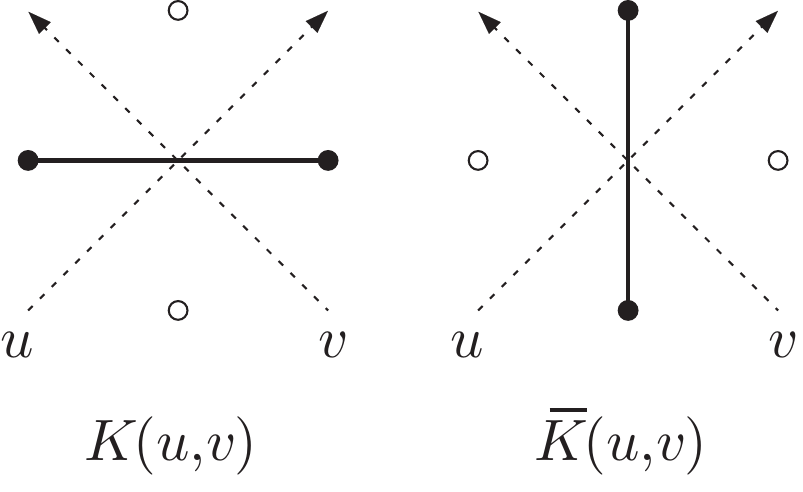}}
\caption{The two kinds of Ising interactions $K$ and $\bar K$. On the
dual lattice $K^{\ast}$ and $\bar K^{\ast}$ are assigned similarly,
but with modulus $k$ replaced by $1/k$.}
\label{fig2}
\end{figure}

The interaction constants $K=\beta J$ are chosen as a function of the two rapidities passing through the bond and the directions of these rapidities, following the prescription of Figure \ref{fig2}, and as a function of the
temperature through the low-temperature elliptic modulus $k$. More
precisely,
\begin{eqnarray}
&&\sinh\big(2K(u,v)\big)=
{\rm sc}(u-v,k')
=k^{-1}{\rm cs}\big({\rm K}(k')-u+v,k'\big),
\label{Ks}\\
&&\sinh\big(2{\bar K}(u,v)\big)
=k^{-1}{\rm cs}(u-v,k')
={\rm sc}\big({\rm K}(k')-u+v,k'\big),
\label{Ksb}
\end{eqnarray}
where ${\rm sc}(v,k)={\rm sn}(v,k)/{\rm cn}(v,k)=1/{\rm cs}(v,k)$.
For the dual lattice with $k^{\ast}=1/k$ being the high-temperature elliptic modulus and
$\sinh(2K^{\ast})\sinh(2{\bar K})=1$, we have
\begin{eqnarray}
&&\sinh\big(2K^{\ast}(u,v)\big)=
k\,{\rm sc}(u-v,k')
={\rm cs}\big({\rm K}(k')-u+v,k'\big),
\label{Ksd}\\
&&\sinh\big(2{\bar K}^{\ast}(u,v)\big)
={\rm cs}(u-v,k')
=k\,{\rm sc}\big({\rm K}(k')-u+v,k'\big).
\label{Ksdb}
\end{eqnarray}
For the triangular lattice we have (\ref{Ks}) with $u-v={\rm K}(k')/3$
or (\ref{Ksb}) with $u-v=2{\rm K}(k')/3$, whereas for the honeycomb
lattice (\ref{Ks}) with $u-v=2{\rm K}(k')/3$ or (\ref{Ksb}) with
$u-v={\rm K}(k')/3$. Therefore, it is easy to see that the resulting
interactions are isotropic for both lattices. 

As the correlation functions only depend on differences of the rapidities,
we can add an arbitrary common constant to all of them \cite{B}. Changing
the direction of a rapidity line is equivalent to adding $\pm{\rm K}(k')$
to its rapidity \cite{APZI}. Together with (\ref{rapidities}), these two
properties show that we have invariance under a rotation by $60^{\circ}$
for the rapidity lattice, implying the required rotation invariance over
$60^{\circ}$ for the pair correlations on the triangular lattice (or over
$120^{\circ}$ for the honeycomb lattice). In addition we have several
reflection properties.

\begin{figure}[tbph]
\centerline{\includegraphics[width=200pt]{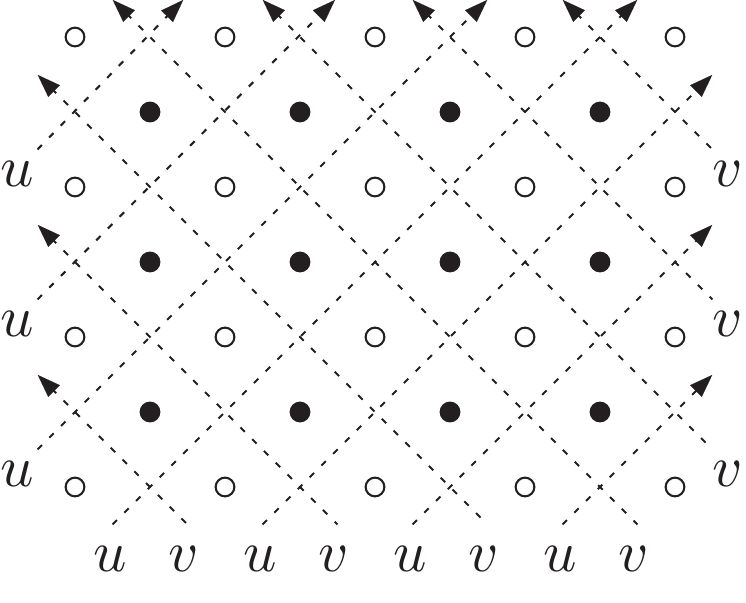}}
\caption{Parts of the infinite square lattice (black circles), dual square
lattice (open circles) and diagonal lattice of rapidity lines (oriented 
dashed lines) with rapidities $u$ and $v$ for the two directions.}
\label{fig3}
\end{figure}

Most importantly, Baxter's $Z$-invariance implies that the pair
correlation functions, apart from their dependence on the modulus $k$,
only depend on the rapidities that pass between the two spins \cite{B},
where we have to make all rapidities pass in the same direction by adding
the above $\pm{\rm K}(k')$ to a rapidity that passes in the opposite
direction \cite{APZI}. Thus we only need to determine universal functions
$g(u_1,\cdots,u_{2m};k)$ and $g^{\ast}(u_1,\cdots,u_{2m};k)$ giving
the pair correlations on the lattice ($T\!<\!T_{\mathrm{c}}$) and the dual lattice ($T\!>\!T_{\mathrm{c}}$).\footnote{Compared to \cite{AP2} we have interchanged $g$ and $g^{\ast}$ through this convention.} These
functions are invariant under any permutation, or under simultaneous
translation by a same amount, of all rapidities \cite{B}. As the
rapidities $u_j$ can only take the three values (\ref{rapidities}), we
find it convenient to introduce the abbreviations \cite{AP1}
\begin{eqnarray}
&&g[N_u,N_v,N_w]\phantom{^{\ast}}\equiv g(u_1,\cdots,u_{2m};k)
\phantom{^{\ast}}=g[N_w,N_v,N_u],\nonumber\\
&&g^{\ast}[N_u,N_v,N_w]\equiv g^{\ast}(u_1,\cdots,u_{2m};k)
=g^{\ast}[N_w,N_v,N_u],
\label{gshort}
\end{eqnarray}
where
\begin{equation}
N_u=\#\{u_j|u_j=u\},\quad N_v=\#\{u_j|u_j=v\},\quad N_w=\#\{u_j|u_j=w\},
\end{equation}
counting the number of $u_j$'s equal $u$, $v$, and $w$. The symmetry
under the interchange of $N_u$ and $N_w$ in (\ref{gshort}) corresponds
to a reflection symmetry that holds in the isotropic case
(\ref{rapidities}). Another reflection symmetry gives
\begin{eqnarray}
&&g[M,N,0]=g[N,M,0]=g[0,M,N]=g[0,N,M],\nonumber\\
&&g[M,0,N]=g[N,0,M],\nonumber\\
&&g[N,0,0]=g[0,N,0]=g[0,0,N],
\label{gother}
\end{eqnarray}
and similar relations hold for $g^{\ast}$; these are also reflection
symmetries for the uniform anisotropic square lattice case represented
in Figure \ref{fig3}.

Now we can invoke the quadratic recurrence relations \cite{P} in the
form \cite{AP2},
\hfill
\begin{eqnarray}
&&{\rm sc}(u_2\!-\!u_1,k'){\rm sc}(u_4\!-\!u_3,k')\nonumber \\
&&\quad\times\big\{g(u_1,u_2,u_3,u_4,\cdots)g(\cdots)\!
-\!g(u_1,u_2,\cdots)g(u_3,u_4,\cdots)\big\}\nonumber \\
&&+\big\{g^\ast(u_1,u_3,\cdots)g^\ast(u_2,u_4,\cdots)\!
-\!g^\ast(u_1,u_4,\cdots)g^\ast(u_2,u_3,\cdots)\big\}=0,\label{Toda1}\\
\cr
&&k^2{\rm sc}(u_2\!-\!u_1,k'){\rm sc}(u_4\!-\!u_3,k')\nonumber \\
&&\quad\times\big\{g^\ast(u_1,u_2,u_3,u_4,\cdots)g^\ast(\cdots)\!
-\!g^\ast(u_1,u_2,\cdots)g^\ast(u_3,u_4,\cdots)\big\}\nonumber \\
&&+\big\{g(u_1,u_3,\cdots)g(u_2,u_4,\cdots)\!
-\!g(u_1,u_4,\cdots)g(u_2,u_3,\cdots)\big\}=0,\label{Toda2}
\end{eqnarray}
where the dots indicate the other rapidities and the modulus that are
left unchanged. Eqs.\ (\ref{Toda1}) and (\ref{Toda2}) are each other's
dual---as is obvious comparing with (\ref{Ks}) and (\ref{Ksd})---and
they can be solved by iteration, once we know the
functions $g$ and $g^\ast$ for the two cases with all or all but one of
the rapidities equal. Such correlation functions are known under the
names diagonal and next-to-the-diagonal correlation functions for the
square-lattice Ising model. An iteration scheme for these is given by
Witte \cite{W}, which we adopt with some modifications.\footnote{In
\cite{W}  a single modulus $k$ with $0<k<\infty$ is used necessitating
definitions such as $\mathrm{K}_<=\mathrm{K}(k)$ and
$\mathrm{E}_<=\mathrm{E}(k)$ for $k<1$ ($T>T_{\mathrm{c}}$) and $\mathrm{K}_>=\mathrm{K}(1/k)$ and
$\mathrm{E}_>=\mathrm{E}(1/k)$ for $k>1$ ($T<T_{\mathrm{c}}$).
Here we opt for two definitions of $k$ to keep $0<k<1$ (cf.\ (\ref{kSaSb})).
Elsewhere we also adopt a single $k$; in (3) and (4) our $k$,
$k_{\mathrm{sq}}$, $k_{\mathrm{tr}}$ and $k_{\mathrm{hc}}$ are to be
identified with $1/k_{\mathrm{Witte}}$ and have been chosen this way because
of our numerical work for which low-temperature expansions have some
advantages.}

Let us introduce the abbreviations\footnote{Here we use the convention of
\cite{MWbook} that in $\sigma_{M,N}$ $M$ is the vertical and $N$ the horizontal coordinate. In many other works, including \cite{W},
the opposite convention is used.}
\begin{equation}
x_n=\langle\sigma_{0,0}\sigma_{n,n}\rangle,\quad
y_n=\langle\sigma_{0,0}\sigma_{n,n+1}\rangle,\quad
z_n=\langle\sigma_{0,0}\sigma_{n+1,n}\rangle,\quad
\end{equation}
for the needed square-lattice correlation functions, together with
\begin{equation}
\mathcal{K}=\frac2{\pi}{\rm K}(k),\quad
\mathcal{E}=\frac2{\pi}{\rm E}(k),\quad
k=\begin{cases}(S_aS_b)^{-1},&T<T_{\mathrm{c}},\cr
S_a^*S_b^*,&T>T_{\mathrm{c}},\end{cases}
\label{kSaSb}
\end{equation}
where
\begin{eqnarray}
&S_a=\sinh(2K_a)=\mathrm{sc}\Big(\frac13\mathrm{K}(k'),k'\Big),\quad
C_a=\cosh(2K_a)=\mathrm{nc}\Big(\frac13\mathrm{K}(k'),k'\Big),&\nonumber\\
&S_b=\sinh(2K_b)=\mathrm{sc}\Big(\frac23\mathrm{K}(k'),k'\Big),\quad
C_b=\cosh(2K_b)=\mathrm{nc}\Big(\frac23\mathrm{K}(k'),k'\Big),&
\label{SaSb}\end{eqnarray}
for $T<T_{\mathrm{c}}$, and
\begin{eqnarray}
&S_a^*=\sinh(2K_a)=\mathrm{cs}\Big(\frac23\mathrm{K}(k'),k'\Big),\quad
C_a^*=\cosh(2K_a)=\mathrm{ns}\Big(\frac23\mathrm{K}(k'),k'\Big),&\nonumber\\
&S_b^*=\sinh(2K_b)=\mathrm{cs}\Big(\frac13\mathrm{K}(k'),k'\Big),\quad
C_b^*=\cosh(2K_b)=\mathrm{ns}\Big(\frac13\mathrm{K}(k'),k'\Big),&
\label{SasSbs}\end{eqnarray}
for $T>T_{\mathrm{c}}$. Eqs.\ (\ref{kSaSb})--(\ref{SasSbs}) define the
rescaled complete elliptic integrals and the hyperbolic sines and cosines
of twice the horizontal and vertical reduced interaction constants in terms
of elliptic modulus $k$. Here $K_a$ is also the reduced interaction energy
$K_{\mathrm{tr}}$ of the triangular lattice and $K_b$ is the
$K_{\mathrm{hc}}$ of the honeycomb lattice.
Duality between low- and high-temperature phases is described
by the replacements
\begin{eqnarray}
&&k^{\ast}=1/k,\quad
\mathcal{K}^{\ast}=k\mathcal{K},\quad
\mathcal{E}^{\ast}=k^{-1}\Big(\mathcal{E}-(1-k^2)\mathcal{K}\Big),
\label{duality1}\\
&&S^{\ast}_a=\frac1{S_b},\quad S^{\ast}_b=\frac1{S_a},\quad
C^{\ast}_a=\frac{C_b}{S_b},\quad C^{\ast}_b=\frac{C_a}{S_a}.
\label{duality2}
\end{eqnarray}
It is easy to check that the dual of the dual gives the original
quantities back ($X^*{}^*=X$).

In general, the nearest-neighbor correlations of the square lattice involve
the complete elliptic integral of the third kind $\Pi_1(n,k)$ \cite{MWbook,APZI,AP1,W},%
\footnote{See (4.3a) and (4.3b) of Chapter 8 of \cite{MWbook}, correcting a
minor misprint, or (52) of \cite{W}, identifying $\Pi_1(n,k)=\Pi(-n,k)$.}
\begin{eqnarray}
&&y_0=\frac2{\pi}\,\frac{C_b}{S_a^{\,2}S\vp_b}\,
\Big[C_a^{\,2}\,\Pi_1(1/S_b^{\,2},k)-\mathrm{K}(k)\Big],\\
&&z_0=\frac2{\pi}\,\frac{C_a}{S\vp_aS_b^{\,2}}\,
\Big[C_b^{\,2}\,\Pi_1(1/S_a^{\,2},k)-\mathrm{K}(k)\Big],
\end{eqnarray}
for $T<T_{\rm c}$, and\footnote{Here, as in (\ref{SasSbs}), the asterisk indicates that the RHS is
the high-temperature expression.}
\begin{eqnarray}
&&y_0^{\ast}=\frac2{\pi}\,\frac{C_b^*}{S_a^*}\,
\Big[C_a^{*2}\,\Pi_1(S_a^{*2},k)-\mathrm{K}(k)\Big],\\
&&z_0^{\ast}=\frac2{\pi}\,\frac{C_a^*}{S_b^*}\,
\Big[C_b^{*2}\,\Pi_1(S_b^{*2},k)-\mathrm{K}(k)\Big],
\end{eqnarray}
for $T>T_{\rm c}$.
However, because we have $K_a=K_{\mathrm{tr}}$, $K_b=K_{\mathrm{hc}}$ and the dual/star-triangle relation (\ref{star-triangle}) or equivalently
\begin{equation}
C_b=\frac{C_a}{C_a-S_a}=C_a(C_a+S_a),
\label{trihon}
\end{equation}
these correlations $y_0$ and $z_0$ (and also $y_0^*$ and $z_0^*$) are also the nearest-neighbour
correlations of the isotropic triangular and honeycomb lattices.
This in turn means they only involve the complete elliptic integral
of the first kind \cite{H,Wan,N}.

To make this more explicit, use \cite{Ha}
\begin{equation}
\Pi_1\left(-k^2\mathrm{sn}^2(a,k),k\right)=\mathrm{K}(k)\left[ 1+
\frac{\mathrm{sn}(a,k)}{\mathrm{cn}(a,k)\mathrm{dn}(a,k)}\,
\mathrm{Z}(a,k)\right],
\end{equation}
where
\begin{equation}
\mathrm{Z}(a,k)=\frac{\Theta'(a,k)}{\Theta(a,k)}
\end{equation}
is Jacobi's Zeta function and \cite{WW}
\begin{eqnarray}
&&\Theta(u,k)=\theta_4(z,q)=
\sum_{n=-\infty}^{\infty} (-1)^n q^{n^2}\mathrm{e}^{2\mathrm{i}nz},
\nonumber\\
&&z=\frac{\pi u}{2\mathrm{K}(k)},\qquad
q\equiv\mathrm{e}^{-\pi\mathrm{K}(k')/\mathrm{K}(k)}.
\label{Theta}
\end{eqnarray}
When $a$ is a rational multiple of $\mathrm{iK}(k')$, say
$a=m\mathrm{iK}(k')/n$, then $\mathrm{Z}(a,k)$ can be expanded
in powers of $q^{1/n}$. It can even be calculated in terms of
$\mathrm{K}(k)$, $S_a$ and $S_b$ using the addition formula \cite{Ha,WW}
\begin{equation}
\mathrm{Z}(u+a,k)=\mathrm{Z}(u,k)+\mathrm{Z}(a,k)-
k^2\,\mathrm{sn}(u,k)\,\mathrm{sn}(a,k)\,\mathrm{sn}(u+a,k),
\label{add}
\end{equation}
and
\begin{equation}
\mathrm{Z}(2\mathrm{iK}(k'),k)=-\frac{\pi\mathrm{i}}{\mathrm{K}(k)},\quad
\mathrm{Z}\big({\textstyle\frac12}\mathrm{iK}(k'),k\big)=
\frac12{\mathrm{i}}(1+k)-\frac{\pi\mathrm{i}}{4\mathrm{K}(k)}.
\end{equation}
Setting $u=2a=4A$, $u=a=2A$ or
$u=a=A\equiv\mathrm{iK}(k')/3$ in (\ref{add}), we find
\begin{eqnarray}
&&\mathrm{Z}(A,k)=
-\frac{\pi\mathrm{i}}{6\mathrm{K}(k)}-
\frac16k^2\mathrm{sn}^3(2A,k)
+\frac12k^2\mathrm{sn}^2(A,k)\,
\mathrm{sn}(2A,k),\nonumber\\
&&\mathrm{Z}(2A,k)=
-\frac{\pi\mathrm{i}}{3\mathrm{K}(k)}-
\frac13k^2\mathrm{sn}^3(2A,k),\qquad
A\equiv\frac13\mathrm{iK}(k').
\end{eqnarray}
Here, using Jacobi's imaginary transformation \cite{WW},
\begin{eqnarray}
&&\mathrm{sn}(A,k)=
\mathrm{i}\,\mathrm{sc}({\textstyle\frac13}\mathrm{iK}(k'),k')=
\mathrm{i}S_a=\frac{\mathrm{i}}{kS_b},\\
&&\mathrm{sn}(2A,k)=
\mathrm{i}\,\mathrm{sc}({\textstyle\frac23}\mathrm{iK}(k'),k')=
\mathrm{i}S_b=\frac{\mathrm{i}}{kS_a}.
\end{eqnarray}
Therefore,
\begin{eqnarray}
&&y_0=\frac13\,\frac{C_a}{S_a}+
\left[\frac{C_b}{S_b}+\frac12\frac{C_a}{S_aS_b}-
\frac16\frac{S_bC_a}{S_a^{\,3}}\right]\mathcal{K},\nonumber\\
&&z_0=\frac23\,\frac{C_b}{S_b}+
\left[\frac{C_a}{S_a}-\frac13\frac{C_b}{S_a^{\,2}}\right]\mathcal{K},
\qquad\mbox{for $T<T_{\rm c}$},
\label{yzlow}
\end{eqnarray}
and
\begin{eqnarray}
&&y_0^{\ast}=\frac13\,C_b+
\left[\frac{C_a}{S_aS_b}+\frac12\frac{C_b}{S_b}-\frac16\frac{S_bC_b}{S_a^{\,2}}\right]\mathcal{K},\nonumber\\
&&z_0^{\ast}=\frac23\,C_a+
\left[\frac{C_b}{S_aS_b}-\frac13\frac{S_bC_a}{S_a^{\,2}}\right]\mathcal{K},
\qquad\mbox{for $T>T_{\rm c}$}.
\label{yzhigh}
\end{eqnarray}
Results (\ref{yzlow}) and (\ref{yzhigh})
differ by duality as defined in (\ref{duality1}) and (\ref{duality2}).

We next rewrite (\ref{yzlow}) and (\ref{yzhigh}) using (\ref{trihon})
or alternatively using $K_a=K_{\mathrm{tr}}$ and $K_b=K_{\mathrm{hc}}$
with the explicit connections to $u$ and $z$ given in
(\ref{k}) and (\ref{star-triangle}).\footnote{Cf.\ also
(\ref{RtimesK}) and (\ref{RoverK}) in the following section.}
With the latter we obtain
\begin{eqnarray}
&&y_0=\frac{1+u}{3(1-u)}\left[1+\frac{2(1-3u)}{\sqrt{(1-u)^3(1+3u)}}
\,\mathcal{K}\right],\\
&&z_0=\frac{1+z^2}{3(1-z^2)}\left[2+\frac{(1+z)(1-4z+z^2)}{(1-z)^3}
\,\mathcal{K}\right],
\end{eqnarray}
which can be compared directly with the internal energy results in
Table I of Houtappel \cite{H}.\footnote{%
The results of Wannier \cite{Wan} and Newell \cite{N}
differ by Landen transformations \cite{WW}
$$
k_{\mathrm{Newell}}=\frac{2\sqrt{k}}{1+k},\quad
k_{\mathrm{Wannier}}=-\frac{1-k'}{1+k'}.
$$} These results are also the basis for
our (\ref{tr1}) and (\ref{h1}); the equality follows by using (\ref{I}) and the low-temperature Landen transformation from (\ref{Landen}) to yield $I(\tau)=2\sqrt{k}\,\mathcal{K}$.

We can also rewrite (44) and (54) of \cite{W}. Then the first few
square-lattice correlations in the low-temperature phase are
\begin{eqnarray}
&&x_0=1,\qquad
x_1=\mathcal{E},\\
&&y_0=
\frac{C_a}{3S_a}\Big(1-
(C_a-2S_a)(C_a+S_a)^2\mathcal{K}^{\ast}\Big),\\
&&z_0=\frac{C_b}{3S_b}\left(2+
\frac{(C_b-2)(C_b+1)^2}{S_b^3}\mathcal{K}\right),\\
&&y_1=\Big(\mathcal{E}-\frac{S_b}{S_a}\mathcal{E}^{\ast}\Big)y_0+
\frac{C_b}{S_a}\mathcal{E}\mathcal{E}^{\ast},\\
&&z_1=\Big(\mathcal{E}-\frac{S_a}{S_b}\mathcal{E}^{\ast}\Big)z_0+
\frac{C_a}{S_b}\mathcal{E}\mathcal{E}^{\ast}.
\end{eqnarray}
whereas the corresponding quantities in the high-temperature phase are
\begin{eqnarray}
&&x^{\ast}_0=1,\qquad
x^{\ast}_1=\mathcal{E}^{\ast},\\
&&y^{\ast}_0=
\frac13C_b\left(1-
\frac{(C_b-2)(C_b+1)^2}{S_b^3}\mathcal{K}\right),\\
&&z^{\ast}_0=\frac13C_a\Big(2+
(C_a-2S_a)(C_a+S_a)^2\mathcal{K}^{\ast}\Big),\\
&&y^{\ast}_1=
\Big(\mathcal{E}^{\ast}-\frac{S_b}{S_a}\mathcal{E}\Big)y^{\ast}_0+
\frac{S_bC_a}{S_a}\mathcal{E}\mathcal{E}^{\ast},\\
&&z^{\ast}_1=
\Big(\mathcal{E}^{\ast}-\frac{S_a}{S_b}\mathcal{E}\Big)z^{\ast}_0+
\frac{S_aC_b}{S_b}\mathcal{E}\mathcal{E}^{\ast}.
\end{eqnarray}
These results are fully consistent with duality defined in (\ref{duality1}) and (\ref{duality2}).
In addition, we have $z^{\ast}_0=C_a-S_ay_0$ and $y^{\ast}_0=C_b-S_bz_0$,
in agreement with (11) in \cite{P}.

Witte's initial conditions, (40) and (42) in \cite{W}, can be
rewritten as
\begin{eqnarray}
&&r_0=1,\qquad\bar r_0=1,\\
&&r_1=-\frac{2k}3+\frac{\mathcal{E}^{\ast}}{3\mathcal{E}},\quad
\bar r_1=\frac{\mathcal{E}^{\ast}}{\mathcal{E}},
\end{eqnarray}
and
\begin{eqnarray}
&&r^{\ast}_0=1,\qquad\bar r^{\ast}_0=1,\\
&&r^{\ast}_1=-\frac2{3k}+\frac{\mathcal{E}}{3\mathcal{E}^{\ast}},\quad
\bar r^{\ast}_1=\frac{\mathcal{E}}{\mathcal{E}^{\ast}}.
\end{eqnarray}
Then further quantities can be found systematically using
\begin{eqnarray}
&&(2j+3)(1-r_j\bar r_j)r_{j+1}
=2j\Big(k+k^{-1}+(2j-1)r_j\bar r_{j-1}\Big)r_j\nonumber\\
&&\qquad\qquad-(2j-3)\Big(1+(2j-1)r_j\bar r_j\Big)r_{j-1},\\
&&(2j+1)(1-r_j\bar r_j)\bar r_{j+1}
=2j\Big(k+k^{-1}-(2j-3)\bar r_jr_{j-1}\Big)\bar r_j\nonumber\\
&&\qquad\qquad-(2j-1)\Big(1-(2j+1)r_j\bar r_j\Big)\bar r_{j-1},
\end{eqnarray}
and the identical equations for $r^{\ast}_j$ and $\bar r^{\ast}_j$,
see (38) and (39) in \cite{W}. The further diagonal and
next-to-the-diagonal correlations follow using
\begin{eqnarray}
&&x_{j+1}=\frac{x_j^2}{x_{j-1}}(1-r_j\bar r_j),\\
&&y_{j+1}=\frac{x_{j+1}}{x_j}
\bigg(1-\frac{\bar r_{j+1}}{\bar r_j}\frac{S_b}{S_a}\bigg)y_j+
\frac{x_{j+1}^2}{x_j^2}\frac{\bar r_{j+1}}{\bar r_j}
\frac{S_b}{S_a}\,y_{j-1},\\
&&z_{j+1}=\frac{x_{j+1}}{x_j}
\bigg(1-\frac{\bar r_{j+1}}{\bar r_j}\frac{S_a}{S_b}\bigg)z_j+
\frac{x_{j+1}^2}{x_j^2}\frac{\bar r_{j+1}}{\bar r_j}
\frac{S_a}{S_b}\,z_{j-1},
\end{eqnarray}
and their dual versions obtained by replacing all quantities by their
$\ast$ versions. These last few equations can be found combining (31),
(36), (59), (63) and (64) of \cite{W}. For the current purpose one
only needs $z\vp_n$ and $z^{\ast}_n$ for $n=0$.

We have now all equations from the square-lattice Ising model
needed to generate $g$ and $g^{\ast}$ with all or all but one of the
rapidities equal in a form that makes the lattice symmetries and duality
manifest. Thus we can now construct a ``polynomial-time" algorithm for
the high- and low-temperature series coefficients for the susceptibility
of the isotropic Ising model on triangular, honeycomb (and kagom\'e)
lattices. For efficiency of the algorithm, we desire series with only
integer coefficients. Series in the low-temperature $u=\exp(-4K_{\mathrm{tr}})$ are certainly acceptable; because the coefficients in these series can be reduced to lattice counts, they are necessarily integer. A useful alternative in the square lattice case \cite{ONGP} was the elliptic parameter $k$. The corresponding alternative here suggested by the $k_{\mathrm{tr}}(u)$ relation (\ref{k}) is an expansion in the variable $\bar k=(k^2/16)^{1/3}$.  Inversion of $k_{\mathrm{tr}}(u)$ results in the series
\begin{equation}
u={\bar k}-2{\bar k}^3+\frac83{\bar k}^4+3{\bar k}^5-16{\bar k}^6+
\frac{152}9{\bar k}^7+40{\bar k}^8-161{\bar k}^9+
\frac{11200}{81}{\bar k}^{10}+\ldots
\label{uink}
\end{equation}
and although the rationals in (\ref{uink}) can be eliminated by the change $\bar k\to\bar k/3$ the coefficients in any correlation function series in $\bar k$ will grow unacceptably rapidly.  A third alternative is expansion in $q^{1/3}$ where $q$ is the elliptic nome.  This is suggested by $\bar k=(k^2/16)^{1/3}$ and the known expansion $k^2/16=q-8q^2+\ldots\,$.

All our elliptic functions
have natural expansions in terms of the elliptic nome
\begin{equation}
q=\exp\left(-\frac{\pi{\rm K}(k')}{{\rm K}(k)}\right),
\end{equation}
using Jacobi theta functions, i.e.\ \cite{WW}
\begin{eqnarray}
&&k=\left[\frac{\theta_2(0,q)}{\theta_3(0,q)}\right]^2,\quad
k'=\left[\frac{\theta_4(0,q)}{\theta_3(0,q)}\right]^2,\\
&&\mathcal{K}=\left[\theta_3(0,q)\right]^2,\quad
\mathcal{E}=\left[\theta_3(0,q)\right]^2-
\frac{\theta_4''(0,q)}{\theta_4(0,q)\left[\theta_3(0,q)\right]^2}.
\end{eqnarray}
Also, from (\ref{SaSb}),
\begin{eqnarray}
&S_a=-\mathrm{i}\;\mathrm{sn}\Big(\frac13\mathrm{i}\;\mathrm{K}(k'),k\Big),\quad
C_a=\mathrm{cn}\Big(\frac13\mathrm{i}\mathrm{K}(k'),k\Big),&\nonumber\\
&S_b=-\mathrm{i}\;\mathrm{sn}\Big(\frac23\mathrm{i}\;\mathrm{K}(k'),k\Big),\quad
C_b=\mathrm{cn}\Big(\frac23\mathrm{i}\mathrm{K}(k'),k\Big),&
\end{eqnarray}	
using Jacobi's imaginary transformation. In terms of theta functions,
\begin{equation}
S_{a,b}=\frac{-\mathrm{i}}{\sqrt{k}}\;
\frac{\theta_1(z_{a,b},q)}{\theta_4(z_{a,b},q)},\qquad
C_{a,b}=\sqrt{\frac{k'}{k}}\;
\frac{\theta_2(z_{a,b},q)}{\theta_4(z_{a,b},q)},
\end{equation}
with
\begin{equation}
z_a=\frac{\pi}{2\mathrm{K}(k)}\;\frac{\mathrm{i}\,\mathrm{K}(k')}3,\quad
z_b=2z_a,\qquad
\mathrm{e}^{\mathrm{i}z_a}=q^{1/6},\quad
\mathrm{e}^{\mathrm{i}z_b}=q^{1/3}.
\end{equation}

From the above we expect to end up with expansions in
the nome
\begin{equation}
\bar q=\exp\left(-\frac{\pi{\rm K}(k')}{3{\rm K}(k)}\right)
=q^{1/3}
\end{equation}
and this is the  good expansion variable that we used.\footnote{There
are many other cases where series in the nome are advantageous. Whenever
all rapidity differences are of the form $m\mathrm{K}(k')/n$ with fixed
integer $n$, we can expand the susceptibility in powers
of $\bar q=q^{1/n}$, see the text following (\ref{Theta}).}

For expansions in terms of the nome it is also advantageous to break
the symmetry defining
\begin{equation}
r_j=(-k)^j\rho_j,\quad \bar r_j=(-k)^{-j}\bar \rho_j,
\end{equation}
and similar for $r^{\ast}_j$ and $\bar r^{\ast}_j$,
in order to avoid square roots of the nome.

\subsection{Alternative expressions \label{bernie}}

The functions $\mathrm{sc}(\frac13\mathrm{K}(k'),k')$ and $\mathrm{sc}(\frac23\mathrm{K}(k'),k')$ have an algebraic representation in $k$ which one can obtain by expanding identities such as
$\mathrm{cs}\big(\textstyle\frac13\mathrm{K}(k)+\frac13\mathrm{K}(k)+\frac13\mathrm{K}(k),k\big)=0$
using standard addition formulae and then solving the resulting quartic equation for $\mathrm{sc}(\frac13\mathrm{K}(k),k)$.  One finds
\begin{equation}
\mathrm{sc}\Big(\frac13\mathrm{K}(k'),k'\Big) = \frac1{\sqrt{Rk}},\quad   \mathrm{sc}\Big(\frac23\mathrm{K}(k'),k'\Big) = \sqrt{\frac Rk},
\label{Rk}
\end{equation}
where
\begin{equation}
R=X+\sqrt{3-X^2+(k^{-1}+k)/X},\quad  X=\sqrt{1+\Big((k^{-1}-k)^2/4\Big)^{1/3}}.
\label{RX}
\end{equation}
Note that $R$ is self-dual, i.e. invariant under the replacement $k\rightarrow1/k$, while $\mathrm{sc}(\frac13\mathrm{K}(k'),k')$ $\leftrightarrow$ $\mathrm{cs}(\frac23\mathrm{K}(k'),k')$. If we take $k= k_{\mathrm{tr}}$, the low-temperature elliptic parameter (\ref{k}), then one can verify
\begin{equation}
\frac{1}{\sqrt{Rk}}=\frac{1-u}{2\sqrt{u}}\equiv\sinh(2K_{\mathrm{tr}})
\label{RtimesK}
\end{equation}
and
\begin{equation}
\sqrt{\frac{R}{k}}=\frac{\sqrt{(1-u)(1+3u)}}{2u}=\frac{1-z^2}{2z}\equiv\sinh(2K_{\mathrm{hc}}),
\label{RoverK}
\end{equation}
where in (\ref{RoverK}) we have used (\ref{star-triangle}) for $u(z)$. In this way we confirm directly from (\ref{Rk})--(\ref{RoverK}) and the definitions (\ref{SaSb}) that $K_a=K_{\mathrm{tr}}$ and $K_b=K_{\mathrm{hc}}$.

The expansions in the (cube root) nome $\bar q=\exp(-\pi\mathrm{K}'/3\mathrm{K})$ described in the preceding section can be applied to $(C_a-S_a)^2$ to give directly $u=u(\bar q)$. We obtain the formula
\begin{eqnarray}
u&=&\bar q\bigg(\sum_{n=0}^{\infty}(\bar q^{\,4n}-\bar q^{\,8n+2})/(1-\bar q^{\,12n+6})\bigg)^2\bigg/\bigg(\sum_{n=0}^{\infty}\bar q^{\,3n(n+1)}\bigg)^4\nonumber\\
&=&\bar q-2\bar q^{\,3}+3\bar q^{\,5}-4\bar q^{\,7}+7\bar q^{\,9}-12\bar q^{\,11}+17\bar q^{\,13}-24\bar q^{\,15}+\dots,
\end{eqnarray}
which explicitly shows $u(\bar q)$ is an integer series. Whether correlation function series in $u$ or $\bar q$
will show the slowest growth in the magnitude of the series coefficients depends on the singularity structure of the correlation functions.  Now the correlation functions as series in $u$ have radius of convergence $1/3$ governed both by the ferromagnetic singularity at $u=1/3$ and an unphysical singularity at $u=-1/3$.  There are other more distant complex singularities and what we have found numerically and describe in section \ref{sec:suppress} is that there is a close analogy with the singularities on the square lattice. Indeed we conjecture that $|k_{\mathrm{tr}}|=1$ is dense with singularities and part\footnote{For the complete natural boundary see Figure \ref{fig5} in Section \ref{sec:suppress}.} of a natural boundary for the triangular lattice. Now the circle $|k_{\mathrm{tr}}|=1$ maps to arcs in the $\bar q$-plane with distance to the origin bounded below by $\exp(-\pi/3)=0.3509\ldots$ and it is this distance that fixes the radius of convergence of the $\bar q$ series. It implies that asymptotically in $N$ we have terms of magnitude $\sim2.85^N\bar q^N$ compared to $\sim3^Nu^N$. As an example of what we observe in practice, the coefficients in the 
series expansion of the low-temperature triangular lattice susceptibility are, at the largest $N$ we have available, dominated by a single $u$-plane singularity pair $\big(u-(1\pm2\mathrm{i})/5\big)^{13/2}$ giving a $\bar q$ series coefficient dependence of magnitude $\sim2.78^N/N^{15/2}$. In conclusion, there is coefficient size reduction in going from series in $u$ to $\bar q$ but it is not dramatic.

\subsection{Computational details\label{sec:computational}}

As discussed in subsection \ref{jacques} the calculation of the triangular and honeycomb lattice susceptibilities as high- and low-temperature series of length $N$ requires as an intermediate step the calculation of two triply indexed arrays $g$ and $g^*$. That is, we are dealing with $\mathrm{O}(N^3)$ elements, each element being a series of length $N$ with integer coefficients whose (digit) size increases linearly with $N$. Fortunately this $\mathrm{O}(N^5)$ memory requirement can be circumvented by a careful sequential arrangement of the calculation and the description of this with emphasis on the storage structure we have implemented is the content of this section. 

\begin{figure}\centering
\includegraphics{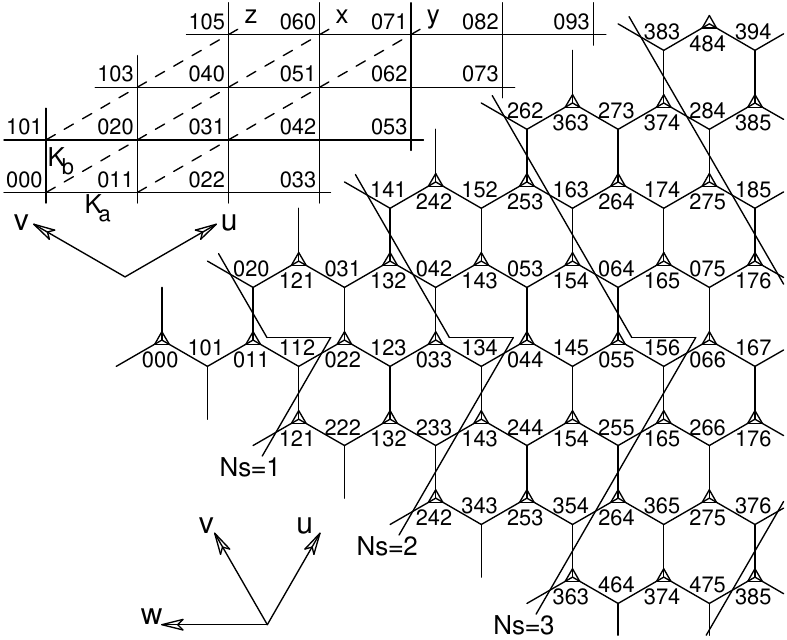}
\caption{A $\pi/3$ section of the triangular/honeycomb lattice. Small triangles mark the triangular and even honeycomb lattice sites. Numbers are the $i,k,j$ indices of the $g$ and $g^{\ast}$ correlation functions. Lines labeled by $N_s$ are the boundaries of the ``shells" described in the text.  The inset shows the $g$ and $g^{\ast}$ labeling of sites on the anisotropic square lattice. Vectors are the directions of the rapidity lines on the two lattices. Dashed lines labeled $x$, $y$ and $z$ indicate the diagonal and near diagonal elements calculated by the Witte recursion relations described in section \ref{jacques}. Note that the two lowest rows on the square correspond exactly to two rows on the triangular lattice.}\label{fig4}
\end{figure}

To begin the discussion we show in Figure \ref{fig4} the $i,k,j$ triples indexing the correlation functions $C(\vec{R})\equiv g(i,k,j)$ at the triangular and honeycomb lattice sites on the minimum sector necessary for obtaining the susceptibility. To obtain these triples refer to Figure \ref{fig1} and let $N_\alpha$ be the number of rapidity lines of type $\alpha$ between the origin and site $\vec{R}$. Then the rules given in Section \ref{jacques} can be summarized by saying that if $\vec{R}$ is in the $\pi/3$ sector above(below) the horizontal through the origin then $g(i,k,j)=g(N_v,N_u,N_w)$ $(=g(N_v,N_w,N_u))$. Clearly if $\vec{R}$ is on the horizontal, $N_v=0$ and $N_u=N_w$. If at least one of $i,k,j$ is zero then the corresponding $g$ is also a correlation function on the anisotropic square lattice. For example, if $i=0$, then in $g(0,k,j)$ we can identify $j=n_x$ and $k=n_y$ where $n_x$ and $n_y$ are the Cartesian coordinates of sites in the first quadrant of the square lattice obtained by rotating that shown in Figure \ref{fig4} clockwise by $\pi/4$. In the fourth quadrant of this rotated lattice $g(i,0,j)=g(-n_y,0,n_x)$. With the exception of $g(1,0,1)$ only elements $g(0,n_y,n_x)$ are needed as initial conditions for the recursion relations for the general $C(\vec{R})$ on the triangular and honeycomb lattices. 

It is worth remarking that the correlations $g(0,n,n)$ on the triangular lattice diagonal can be calculated as Toeplitz determinants \cite{Stephenson} and we have used this as an important check of our computations. We also note that of the symmetries (\ref{gshort}) and (\ref{gother}) satisfied by the $g(i,k,j)$, two in particular that we use below are $g(i,k,j)=g(j,k,i)$ and $g(0,k,j)=g(0,j,k)$.

One observes in Figure \ref{fig4} that within each ``shell" $N_s$, that is, sites between lines $N_s-1$ and $N_s$, the central $k$ index is either $2N_s-2$ or $2N_s-1$. It turns out that with a few exceptions, an array at fixed $k$ can be computed from elements in an array with index $k-1$. Proceeding sequentially through ``shells", or equivalently $k$, reduces the memory requirement from $\mathrm{O}(N^5)$ to $\mathrm{O}(N^4)$. It also has the advantage of allowing the calculation to be stopped and restarted if necessary at convenient intervals and makes calculation with an $N$ of several hundred to a thousand practical. 

We take the array\footnote{While we only refer to an array $g$ here, there is a strictly parallel dual array $g^*$. It is to be understood that such convention applies throughout this section.} $g(i,k,j) \equiv g_k(i,j)$ indexed---using Maple notation---in the double sequence
\texttt{seq(seq(gk(i,i+2*j),j=0\ldots Ns-i),i=0\ldots Ns)} for even $k=2 N_s-2$ as constituting ``shell" $N_s(a)$. The array $g(i,k,j) \equiv g_k(i,j)$ with odd $k=2 N_s-1$ and indexed as \texttt{seq(seq(gk(i,i+2*j+1),j=0\ldots Ns-i),i=0\ldots Ns)} constitutes ``shell" $N_s(b)$. The indexing for both arrays is such that $j \geq i, i+j \leq k+2$, and satisfies the requirement that $i+k+j$ be even. As a specific example of this indexing, the $N_s=3$ case is illustrated as (\ref{40})
\begin{align}\label{40}
(g_4) \hspace{0.5cm} && 040 && \overbrace{042}^\uparrow && \overbrace{044}^\Uparrow && 046 && \overbrace{141}^\uparrow && \overbrace{143}^\Uparrow && \overbrace{145}^\uparrow && \overbrace{242}^\Uparrow && \overbrace{244}^\uparrow && \overbrace{343}^\uparrow && \hspace{0.5cm} (3a) \nonumber \\
(g_5) \hspace{0.5cm} && 051 && \overbrace{053}^\uparrow && \overbrace{055}^\Uparrow && 057 && \overbrace{152}^\uparrow && \overbrace{154}^\Uparrow && \overbrace{156}^\uparrow && \overbrace{253}^\Uparrow && \overbrace{255}^\uparrow && \overbrace{354}^\uparrow && \hspace{0.5cm} (3b) 
\end{align}
with $g_k$ label on the left and ``shell" label on the right. Both $N_s(a)$ and $N_s(b)$ arrays are of length $L=(N_s+1)(N_s+2)/2$ and we introduce a third notation, namely the single indexed $g_k(\ell), \ell=1\ldots L$. The physical elements on the lattice are only a subset of $3N_s-1$ elements in each array. Specifically, the triangular and even honeycomb sites are at locations $\ell=L-1-n(n+1)/2, n=1\ldots N_s$ and are indicated by the double arrows in (\ref{40}). The odd honeycomb sites below the horizontal in Figure \ref{fig4} are at $\ell=L-n(n+1)/2, n=0\ldots N_s-1$ while those above are at $\ell=L-2-n(n+1)/2, n=2\ldots N_s$. Both sets are indicated by single arrows in (\ref{40}).

We also require linear arrays which for identification purposes we will denote as $d_k$ with the even and odd $k$ arrays being distinct. The array $d_0$ is initialized by elements from the anisotropic square lattice array $x$ described in Section \ref{jacques}; $d_1$ by the corresponding elements from $y$. In subsequent calculations, $d_{k-2}$ will be renamed $d_k$ and certain elements changed by an in-place replacement determined by the quadratic recursion formulae. Details will be described below; for now it is enough to know that the changes will maintain $d_k(1)=g(0,k+2,k \mod 2)$, $d_k(2)=g(0,k,k+2)$ and $d_k(3)=g(0,k,k+4$). The $d_k(n), n>N_s+1$ remain unchanged from the initializations 
\begin{align}\label{42}
(d_0) && \uline{020} && \uline{002} && \uline{004} && \uline{006} && \uline{008} && \uline{\ldots} && \hspace{1.5cm} (g_0) && \overbrace{\uline{000}}^\Uparrow && \uline{002} && \overbrace{\uline{101}}^\uparrow && (1a) \nonumber \\
(d_1) && \uline{031} && \uline{013} && \uline{015} && \uline{017} && \uline{019} && \uline{\ldots} && \hspace{1.5cm} (g_1) && \overbrace{\uline{011}}^\Uparrow && \uline{013} && \overbrace{112}^\uparrow && (1b) 
\end{align}
where arrows indicate physical site elements as in (\ref{40}). The underlines indicate elements that have been copied, specifically $d_0(n)=x_{n-1}$ and $d_1(n)=y_{n-1}$ for $n>1$. Also, the elements in $g_0$ are $x_0$, $x_1$ and $z_0$ while the first two in $g_1$ are $y_0$ and $y_1$. A special remark is in order for elements $d_0(1)$ and $d_1(1)$---these are equal respectively to $d_0(2)$ and $d_1(2)$ because of the symmetry $g_k(0,j)=g_j(0,k)$. The third element in $g_1$ is given by 
\begin{equation}\label{43}
g(1,1,2)=g(0,1,1)g(1,0,1)+g^*(0,1,1)(g^*(1,0,1)-g^*(0,0,2))k_{\mathrm{tr}}
\end{equation}
which is a special case of the recursion equation (\ref{48}).
Note that the dual of (\ref{43}) requires both $g \leftrightarrow g^*$ and $k_{\mathrm{tr}} \rightarrow 1/k_{\mathrm{tr}}$.

This completes the initialization except for combining the physical site elements in (\ref{42}), with appropriate multiplicity factors, into (summed) correlation functions from which susceptibilities will be determined as a very last step. These functions are chosen to distinguish between even and odd sites; given the initialization (\ref{42}) we set 
\begin{align}\label{44}
C_\mathrm{e}& = \delta g_0(1) + 6 \delta g_1(1), & C_\mathrm{o} = & 3 \delta g_0(3) + 6 \delta g_1(3), \nonumber \\
C_\mathrm{e}^* & = g_0^*(1) + 6 \delta g_1^*(1), & C_\mathrm{o}^* = & 3 g_0^*(3) + 6 g_1^*(3).
\end{align}
Each $\delta g$ in (\ref{44}) is the magnetization subtracted $g-M^2$ which applies only to the low-temperature variables and not the high-temperature duals. 

The recursion in which new $g_k$ are calculated starts with $k=2$ and $N_s=2$. In the general case the first element of $g_k$ is initialized by copying from $d_{k-2}$, specifically $g_k(1)=d_{k-2}(1)=g(0,k,k \mod 2)$.  We then proceed sequentially from the $g_k(2)$ to the final $g_k(L), L=(N_s+1)(N_s+2)/2$, using the quadratic recursion relations for each. Unless forced otherwise, we use only elements from $g_k$ and $g_{k-1}$ to minimize what is kept in memory and this requires that different forms of the recursion relations be used depending on the $i,j$ combination in $g_k(i,j)$. In the order used, these are\footnote{As noted in the context of (\ref{40}) each equation is to be understood as a pair. Here the second member is obtained by the interchange $g_k \leftrightarrow g_k^*$ and replacements $d^* \rightarrow d$ and $k_{\mathrm{tr}} \rightarrow 1/k_{\mathrm{tr}}$.} 
\begin{eqnarray}
g_k(0,j) & = & [g_{k-1}(0,j-1)^2+(g_{k-1}^*(0,j-1)^2 \nonumber \\
 & & -g_k^*(0,j-2)g_{k-2}^*(0,j))R k_{\mathrm{tr}}]/g_{k-2}(0,j-2), \hspace{1cm} 1<j\leq k, \label{45}\\
g_k(0,k+2) & = & [g_{k-1}(0,k+1)^2+(g_{k-1}^*(0,k+1)^2 \nonumber \\
 & & -g_k^*(0,k) d_{k-2}^*(3))R k_{\mathrm{tr}}]/g_{k-2}(0,k), \label{46} \\
g_k(1,1) & = & [g_{k-1}(0,1)^2-(g_{k-1}^*(0,1)^2 \nonumber \\
 & & -g_k^*(0,0)g_{k-2}^*(1,1))R k_{\mathrm{tr}}]/g_{k-2}(0,0) \label{47}\\
g_k(1,j) & = & [g_k(0,j-1)g_{k-1}(1,j-1) \nonumber \\
 & & +(g_k^*(0,j-1)g_{k-1}^*(1,j-1)-g_k^*(1,j-2)g_{k-1}^*(0,j))k_{\mathrm{tr}}] \nonumber \\
 & & /g_{k-1}(0,j-2), \hspace{1cm} j>1 \label{48} \\
g_k(i,j) & = & [g_k(i-1,j-1)g_{k-1}(i-1,j) \nonumber \\
 & & +(g_k^*(i-1,j-1)g_{k-1}^*(i-1,j)-g_k^*(i-2,j)g_{k-1}^*(i,j-1))k_{\mathrm{tr}}] \nonumber \\
 & & /g_{k-1}(i-2,j-1), \hspace{1cm} i \geq 2 \label{49}
\end{eqnarray}
where the (self-dual) multiplier $R$ is given in (\ref{RX}) or, more simply, as $R=S_b/S_a$ by combining (\ref{RtimesK}) and (\ref{RoverK}).
The $j$ index in these recursion equations increments in steps of two to maintain $i+j+k$ even, a condition that also eliminates (\ref{47}) unless $k$ is even. Indexing functions are easily established which relate the location of the right hand side elements in (\ref{45})--(\ref{49}) to those on the left; this is a coding detail that we do not give here except to remark that the symmetry $g_k(i,j)=g_k(j,i)$ may have to be invoked to locate an element. The special element $d_{k-2}^*(3)$ in (\ref{46}) is $g^*(0,k-2,k+2)$ which in our construction of the $g_{k-2}$ array was explicitly excluded from being one of the elements. As an observation on memory requirements, only the first $N_s$ elements of array $g_{k-2}$ are required for implementing (\ref{45})--(\ref{47}) so that most of the memory used by $g_{k-2}$ could be released before the $g_k$ calculation is started. For all further calculations in (\ref{48}) and (\ref{49}) only $g_{k-1}$ need be maintained in memory. In fact with a small location offset of $2N_s+1$ the replacement $g_{k-1} \rightarrow g_k$ could be done in-place and thus reduce memory requirements even further. On completion of the $g_k$ calculation in (\ref{45})--(\ref{49}) the $g_k$ elements corresponding to physical lattice sites are accumulated into the $C$ and $C^*$ as in (\ref{44}) with appropriate attention to multiplicity. 

We must also update the $d_{k-2}$ array that has just been used in (\ref{46}) in preparation for subsequent iterations in $k$. The $d_k$, and for completeness the relevant $g_k$, are shown in (\ref{51})
{\footnotesize\begin{align}
(d_0) && \uline{020} && \uline{002} && \uline{004} && \uline{006} && \uline{008} && \uline{\ldots} && && && && && && (g_0) && \overbrace{\uline{000}}^\Uparrow && \uline{002} && \overbrace{\uuline{101}}^\uparrow && (1a) \nonumber \\
(d_1) && \uline{031} && \uline{013} && \uline{015} && \uline{017} && \uline{019} && \uline{\ldots} && && && && && && (g_1) && \overbrace{\uline{011}}^\Uparrow && \uline{013} && \overbrace{112}^\uparrow && (1b) \nonumber \\
&& && && && && && && && && (g_2) && \overbrace{\uline{020}}^\uparrow && \overbrace{022}^\Uparrow && 024 && \overbrace{121}^\Uparrow && \overbrace{123}^\uparrow && \overbrace{222}^\uparrow && (2a) \nonumber \\
(d_2) && \uline{040} && \uline{024} && 026 && 006 && \ldots \nonumber \\
&& && && && && && && && && (g_3) && \overbrace{\uline{031}}^\uparrow && \overbrace{033}^\Uparrow && 035 && \overbrace{132}^\Uparrow && \overbrace{134}^\uparrow && \overbrace{233}^\uparrow && (2b) \nonumber \\
(d_3) && \uline{051} && \uline{035} && 037 && 017 && \ldots \nonumber \\
&& && && && && (g_4) && \uline{040} && \overbrace{042}^\uparrow && \overbrace{044}^\Uparrow && 046 && \overbrace{141}^\uparrow && \overbrace{143}^\Uparrow && \overbrace{145}^\uparrow && \overbrace{242}^\Uparrow && \overbrace{244}^\uparrow && \overbrace{343}^\uparrow && (3a) \nonumber \\
(d_4) && \uline{060} && \uline{046} && 048 && 028 && 008 && \ldots \nonumber \\
&& && && && && (g_5) && \uline{051} && \overbrace{053}^\uparrow && \overbrace{055}^\Uparrow && 057 && \overbrace{152}^\uparrow && \overbrace{154}^\Uparrow && \overbrace{156}^\uparrow && \overbrace{253}^\Uparrow && \overbrace{255}^\uparrow && \overbrace{354}^\uparrow && (3b) \nonumber \\
(d_5) && \uline{071} && \uline{057} && 059 && 039 && 019 && \ldots \nonumber \\
(g_6) && \uline{060} && 062 && \overbrace{064}^\uparrow && \overbrace{066}^\Uparrow && 068 && 161 && \overbrace{163}^\uparrow && \overbrace{165}^\Uparrow && \overbrace{167}^\uparrow && \overbrace{262}^\uparrow && \overbrace{264}^\Uparrow && \overbrace{266}^\uparrow && \overbrace{363}^\Uparrow && \overbrace{365}^\uparrow && \overbrace{464}^\uparrow && (4a) \nonumber
\end{align}}\begin{equation}\label{51}\end{equation}
to illustrate the changes in the $d_k$ as one proceeds through to the completion of ``shell" 3 and into ``shell" 4. The notation in (\ref{51}) is as in (\ref{40}) and (\ref{42}). Of special note are the underlined elements in $d_k, k\geq2$, and the fact that all changes are made in-place. Specifically this means that we first rename $d_{k-2}$ to $d_k$. Then we copy $d_k(N_s+1)$ to $d_k(1)$ since it is needed both in its original location where it will be overwritten and in a subsequent $g_{k+2}$ calculation.\footnote{There are in general other elements that could be saved for $g_{k+2m}, m>1$, but we have opted instead for a small amount of redundancy in our calculation.} The second copy is from the just completed $g_k(N_s+1)$ to $d_k(2)$. The transformation of $d_k$ is then completed by a sequence of in-place quadratic recursion transformations of elements $d_k(n)$ starting at $n=N_s+1$ and decrementing to $n=3$. Each recursion is given by\footnote{Once again there is a second member obtained by $d \leftrightarrow d^*$ and $k_{\mathrm{tr}} \rightarrow 1/k_{\mathrm{tr}}$ which must be done before $n$ is decremented.} 
\begin{equation}
d_k(n) = [d_{k-1}(n)^2+(d_{k-1}^*(n)^2-d_k^*(n-1)d_k^*(n+1))R k_{\mathrm{tr}}]/d_k(n)
\end{equation}
which is in a form identical to (\ref{45}) including $R$ from (\ref{RX})--(\ref{RoverK}). All operations in ``loop" $k$ have now been completed and we can restart the overall cycle begun following (\ref{44}) after incrementing $k \rightarrow k+1$ and, if the new $k$ is even, $N_s \rightarrow N_s+1$.

On completion of all recursions, high- and low-temperature susceptibility series are generated from the $C$ and $C^*$ as follows. The triangular lattice susceptibility for $T<T_{\mathrm{c}}$ is given directly as 
\begin{equation}
k_{\mathrm{B}}T\,\chi_-^{\mathrm{tr}}(u) = C_\mathrm{e}(u)
\end{equation}
while that for the honeycomb follows from the duality/star-triangle transformation (\ref{star-triangle}) and is 
\begin{equation}\label{54}
k_{\mathrm{B}}T\,\chi_-^{\mathrm{hc}}(z) = C_\mathrm{e}\left(u = z/(1-z+z^2)\right) \pm C_\mathrm{o} \left( u = z/(1-z+z^2) \right).
\end{equation}
Note that both odd and even sites contribute in (\ref{54}) with the sum for the ferromagnet; the difference for the antiferromagnet. The results for $T>T_{\mathrm{c}}$ follow by duality and are 
\begin{eqnarray}
k_{\mathrm{B}}T\,\chi_+^{\mathrm{tr}}(v) & = & C_\mathrm{e}^*\left(u = v/(1-v+v^2) \right), \\
k_{\mathrm{B}}T\,\chi_+^{\mathrm{hc}}(v) & = & C_\mathrm{e}^*\left(u = v^2 \right) + C_\mathrm{o}^* \left(u = v^2\right),
\end{eqnarray}
where $v=\tanh(K)$ is the conventional high-temperature variable and $K$
is $K_{\mathrm{tr}}$ or $K_{\mathrm{hc}}$ as appropriate.
All of these susceptibilities agree with the earlier work by Sykes et al.\ \cite{Sykes1, Sykes2, Sykes3}. 

In our implementation of the above procedure we made full use of Maple's automatic series multiplication routines in full integer arithmetic. This is similar to what was done in \cite{ONGP} for the square lattice and allowed us to reach series of adequate length. However we did introduce several modifications to improve efficiency. First, as also in \cite{ONGP}, when generating high- and low-temperature series the recursions were set up to deal directly with the much smaller residuals $\delta g=g-M^2$. As an example of this change, the recursion (\ref{49}) becomes
\begin{eqnarray}
\delta g_k(i,j) & = & \delta g_k(i-1,j-1)+\delta g_{k-1}(i-1,j)-\delta g_{k-1}(i-2,j-1) \nonumber \\
 & & \big[\big(\delta g_k(i-1,j-1)-\delta g_{k-1}(i-2,j-1)\big) \nonumber \\
 & &\quad\times\big(\delta g_{k-1}(i-1,j)- \delta g_{k-1}(i-2,j-1)\big) \nonumber \\
 & & +\big(g_k^*(i-1,j-1)g_{k-1}^*(i-1,j)-g_k^*(i-2,j)g_{k-1}^*(i,j-1)\big)k_{\mathrm{tr}}\big] \nonumber \\
 & & /\big(M^2+\delta g_{k-1}(i-2,j-1)\big), \hspace{1cm} i \geq 2
\end{eqnarray}
in which the magnetization appears only in a denominator factor. 

A second change was based on the observation that all $g^*$ terms on odd honeycomb sites are of the form $\sqrt{u}$ times series in $u$. If we define these $g^*$ terms as $\tilde{g}^* k_{\mathrm{tr}}$ and use $\tilde{g}^*$ in the recursion relations in place of $g^*$ one can eliminate all occurrences of $\sqrt{u}$ and dramatically speed up Maple's handling of the resulting series\footnote{The rescaling also has the advantage of eliminating about
one-half of all explicit occurrences of $k_{\mathrm{tr}}$ in the recursions
(\ref{45})--(\ref{49}) and thus reducing the number of required series
multiplications}. 

Thirdly, we transformed from series in $u$ to series in the (cube root) nome $\bar q=\mathrm{e}^{-\pi\mathrm{K}'/3\mathrm{K}}$.  As remarked in Section \ref{bernie}, the effect is not dramatic but because the implementation of a variable change is so easy we did take this opportunity for improved efficiency.

For the high- and low-temperature susceptibilities, we generated series to ``shell" 160 in about 40 days on a 3 Ghz Pentium processor with 500 Mbyte memory. This gives $\chi^{\mathrm{tr}}(u)$, $\chi^{\mathrm{hc}}(v)$ and $\chi^{\mathrm{hc}}(z)$ to about 640 terms and $\chi^{\mathrm{tr}}(v)$ to about 320 and these series can be found in \cite{Sourcefiles}. 

We have also run the recursion program for series in $\tau$ to $\mathrm{O}(\tau^{23})$ for the data necessary to determine the ``short-distance" terms in $\chi$. Here there is no magnetization subtraction; instead $g^*(\tau)=g(-\tau)$ and the code simplifies considerably. It is only practical to run in floating point and we have gone as high as 121 ``shells" with an accuracy estimated better than about 500 digits. Another difference from the high- and low-temperature series case is that the correlation data from different shells is not accumulated but rather kept separate so as to allow a fitting procedure completely analogous\footnote{One important observation is that the factor $\sqrt{s}$ that appears in various equations in \cite{ONGP} is now to be interpreted as $k^{1/4}$---it remains as the same function of $\tau$.} to that described in \cite[Section 6]{ONGP}. Note that there is a distinction between what constitutes a shell for short-distance fitting and the ``shells" as defined in Figure \ref{fig4}. First, a fitting shell contains only one layer each of odd and even sites---not the two shown in Figure \ref{fig4}---so our data extends to 241 fitting shells. Secondly, we try to keep fitting shells as close to perfect hexagons as possible. Symmetry dictates that we use even sites from a single $g_k$ but odd sites are taken from $g_k$ if they are below the horizontal in Figure \ref{fig4} and from $g_{k+1}$ if they are above the horizontal. 

The individual shell values are of little intrinsic interest and are not recorded here. Instead we give ``short-distance" terms, which are the output of the fitting procedure, in an abbreviated form in Appendix C and to the full estimated accuracy of our calculations in \cite{Sourcefiles}. To complement the much longer 2042 term high- and low-temperature square lattice series from \cite{Bou}, we have rerun the code in \cite{ONGP} for series in $\tau$ to $\mathrm{O}(\tau^{29})$ to 241 shells. Our extended fits confirm the earlier results from \cite{ONGP} and the new output is recorded in Appendix C and \cite{Sourcefiles} as for the triangular/honeycomb data. 

\section{Extracting the scaling function\label{sec:extract}}

\subsection{Changing the series variable\label{sec:convert}}

Once we  obtained the high- and low-temperature susceptibility series, we analysed them to extract the scaling function. Firstly, we normalised the series variable so that the ferromagnetic singularity occurs at 1. For example, for the high- and low-temperature square lattice series we use the variables $z = s$ and $z = 1/s^2$ respectively.

We began with short-distance terms calculated from the expansion of the susceptibility in terms of $\tau$ as described in \cite[Section 6]{ONGP}, and a number of Aharony and Fisher scaling terms which are known to be accurate. These are of the form $\tau^a (\ln |\tau|)^b$ and $\tau^{-7/4+a}$ respectively.

We converted these to series in our chosen variable $z$ in the following manner. First we expressed each of these terms as a series in $1-z$, of order approximately 50, which may be multiplied by $(\ln (1-z))^b$ or $(1-z)^{-7/4}$. Each term in the $1-z$ series was then expanded as a series in $z$ to the full length of the susceptibility series (about 2000 terms for the square lattice, for example), and the results added up to produce a series in $z$ for each short-distance and Aharony and Fisher scaling term. All these series were then subtracted from the susceptibility series. This formed a new series $\sum_n c_n z^n$, singular at $z = 1$.

\subsection{Singularity suppression}\label{sec:suppress}

The next step involved suppressing the effect of the competing singularities on the series. For the square lattice, the singularities of the susceptibility are given (\cite{Nic1}) by the singularities of the $N$-particle contributions. These lie on the unit circle $|s| = 1$ at the points $s_{kl} = \exp(i \theta)$, where
\begin{equation}
2 \cos \theta = \cos \frac{2 \pi k}{N} + \cos \frac{2 \pi l}{N}, \hspace{1cm} 0 \leq k,l < N \mbox{ ($k$,$l$ not both 0)}.
\end{equation}

For the low-temperature series, only the even-$N$ singularities are relevant. The asymptotic behaviour of the susceptibility near each of these singularities is given (\cite{Nic2}) by $(1 - z/z')^p$, where $z'$ is the singularity and $p = (N^2-3)/2$. This introduced a term into the susceptibility series which behaves asymptotically as $n^{-p-1}$ (since $|z'| = 1$). This has the potential to dominate the effect of the scaling term $\tau^{-7/4 + a} \sim (1 - z)^{-7/4+a}$ for large $a$, since it introduces a term into the susceptibility which behaves asymptotically as $n^{3/4-a}$.

The simplest procedure (which was the one used in \cite{ONGP}) to rectify this is simply to multiply the series by $1 - z/z'$. This changes the behaviour of the contribution from the singularity at $z'$ to $n^{-p-2}$, but leaves the contribution from the scaling term at $n^{3/4-a}$.

However, because we know the exact form of the singularity, we can use a more accurate suppression. To illustrate, we begin by observing that
\begin{equation}
I_p \left(1 - \frac{z}{z'} \right)^p \equiv \left[ \left(1 - \frac{z}{z'}\right) + \frac{p+1}{z'} \int dz \right] \left(1 - \frac{z}{z'}\right)^p = c,
\end{equation}
where $c$ is a constant. Expressing the original singularity term as a series shows that applying $I_p$ forms the new series
\begin{equation}\label{eq:suppress}
\sum_{n} \left[ c_n - \frac{1}{z'} \left(1 - \frac{p + 1}{n} \right) c_{n-1} \right] z^n
\end{equation}
which completely removes the $(1 - z/z')^p$ term. Moreover, because
\begin{equation}
I_p \left(1 - \frac{z}{z'}\right)^{p+a} = -\frac{a}{(p+a+1)} \left(1 - \frac{z}{z'}\right)^{p+a+1} + c,
\end{equation}
this transformation also has the additional effect of suppressing $(1 - z/z')^{p+1}$. In other words, the contribution to the susceptibility from this singularity goes from $n^{-p-1}$ to $n^{-p-3}$ when we apply this suppression, compared with $n^{-p-2}$ when we simply multiply by $1 - z/z'$.

In addition, applying the integral operator to scaling terms gives
\begin{equation}
I_p \left(1 - z\right)^{-7/4+a} = \left(1 - \frac{z}{z'}\right) (1-z)^{-7/4+a} - \frac{p+1}{z'(-7/4+a+1)} (1-z)^{-3/4+a},
\end{equation}
which still contributes $n^{3/4-a}$ to the asymptotic behaviour of the susceptibility series. So this operator suppresses the competing singularity while not asymptotically affecting the scaling term.

An unfortunate consequence of applying $I_p$ for a complex singularity is that the series resulting from (\ref{eq:suppress}) has complex coefficients. This can be avoided by observing that since the susceptibility is real, for every singularity $z'$ there is a corresponding singularity of the same order at $\bar{z'}$. Sequentially applying the suppression to both of these singularities results in the series
\begin{equation}
\sum_{n} \left[ c_n - \frac{2 \mbox{ Re } z'}{|z'|^2} \left(1 - \frac{p + 1}{n} \right) c_{n-1} + \frac{1}{|z'|^2} \left(1 - \frac{p+1}{n}\right)\left(1 - \frac{p+1}{n-1}\right) c_{n-2}\right] z^n,
\end{equation}
which can be seen to have real coefficients. As all we are doing is applying formula (\ref{eq:suppress}) twice for two different singularities, the effects on the singularity and scaling terms that we observed above still hold.

In practice, we also suppress the higher-order terms $(1 - z/z')^{p+a}$ for $a = 2, 4, \ldots$, using the above suppression formula (with $p$ replaced by $p+a$) for each $a$. The maximum $a$ that we use varies for each singularity and is determined empirically as described below.

For high-temperature series, only the odd-$N$ singularities are relevant. The asymptotic behaviour of the susceptibility near each of these singularities is given (\cite{Nic1}) by $(1 - z/z')^p \ln (1 - z/z')$, where $p = (N^2-3)/2$. These terms can also be suppressed by the same formula (\ref{eq:suppress}). This can be seen to be true because applying the same integral operator results in an analytic term for integer $p$ (which is true for odd $N$). Again, we suppress a number of higher powers.

In order to determine which singularities should be suppressed and by how much, we apply a Fast Fourier Transform diagnostic, as described in \cite[Section 7]{Bou}. We first do a preliminary fit of the series to our functions, as described in section \ref{sec:fit} below, and subtract the fit from the series. The dominant unsuppressed singularity in the remainder is expressed by periodic behaviour of period $2 \pi/\theta$, for a singularity located at $\exp(i\theta)$. By applying FFT to the remainder, we can observe the periods of the dominant unsuppressed singularities, match these to the known singularities, and increase the suppression on these singularities (by suppressing more higher-order terms). This is repeated until the remainder has a satisfactorily small amplitude.

The analysis of the triangular and honeycomb series is almost identical, though we must suppress the appropriate singularities (see \cite{Vai}). For these lattices, it is conjectured that the singularities lie on the curve of Matveev and Shrock (\cite{Mat}),
\begin{equation}\label{36}
1 + 3 u^2 - 2 u (1-u) x = 0, \hspace{1cm} -\frac{3}{2} \leq x \leq 3.
\end{equation}
We further conjecture that the singularities are given implicitly by this equation when $x$ takes the values
\begin{equation}\label{37}
x_{klm} = \cos \frac{2 \pi k}{N} + \cos \frac{2 \pi l}{N} + \cos \frac{2 \pi m}{N}, \,\,\, k + l + m \equiv 0 \,\, {\rm mod} \, N \mbox{ ($k$,$l$,$m$ not all 0)}.
\end{equation}
To suppress these singularities, we again apply formula (\ref{eq:suppress}), assuming that the form of the singularities, and in particular the exponent $p = (N^2-3)/2$, is the same for these lattices as for the square lattice.

Partial confirmation of this conjecture arises from the singularities that we observe from our FFT diagnostic as we suppress singularities. We have observed the singularities corresponding to this formula for $(k,l,m) = (1,0,-1)$ for $N = 3$ to 8 and $N = 10$, and for $(k,l,m) = (2,0,-2), (2,-1,-1)$ for $N = 6$.

\begin{figure}\centering
\includegraphics{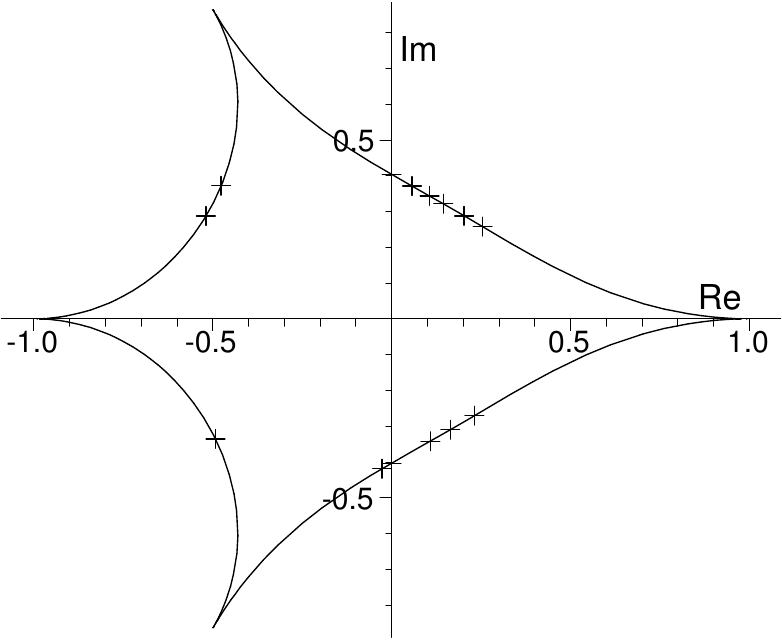}
\caption{The conjectured natural boundary in the complex (cube root) nome $\bar{q}$-plane for the Ising model on the triangular lattice. The real axis cusps are the points $u=\pm1/3$; the other two are $u=-1+\mathrm{i}0^\pm$. The right side arcs are defined by $|k_{\mathrm{tr}}|=1$ and $u=(-1+2\mathrm{e}^{\mathrm{i}\phi})/3, -\pi<\phi<\pi$. The left arcs correspond to straight line segments lying on either side of the cuts, $-\infty<k_{\mathrm{tr}}\leq-1$ in $k_{\mathrm{tr}}$ and $-1\leq u\leq-1/3$ in $u$. Crosses mark the singularities found in the series analysis described in the text. For clarity, the singularities for $T<T_{\mathrm{c}}$ are shown only in the upper half plane, those for $T>T_{\mathrm{c}}$ in the lower half.}\label{fig5}
\end{figure}

We checked for additional singularities by analyzing both low- and high-temperature series of the triangular lattice susceptibility in the (cube root) nome $\bar{q}=\exp(-\pi K'/3K)$. Because all complex portions of the $\bar{q}$-plane curves defined by (\ref{36}) are at or within the distance to the ferromagnetic singularity, the high order series coefficients in $\bar{q}$ will be dominated by the complex singularities. By a succession of suppressions of the dominant terms and FFT diagnostics we have identified the same $N=4, 6$ and 8 singularities as found in the $u$-plane; in addition $(k,l,m)=(2,-1,-1)$ for $N=8$ and two singularities consistent with $x \approx -1.21$ and -1.35 in (\ref{36}). The latter singularities are those on the left, upper plane arc shown in Figure \ref{fig5}. From the $T>T_{\mathrm{c}}$ series in $\bar{q}$ we find the same $N=3, 5$ and 7 as in the $v$-plane, the $(k,l,m)=(2,-1,-1)$ for both $N=5$ and 7 and a singularity consistent with $x \approx -1.27$ in (\ref{36}) and shown on the left, lower arc in Figure \ref{fig5}. The singularities on the left arcs are not identifiable with any small integer values in (\ref{37}). Thus, although we propose that the closed curve in Figure \ref{fig5} is a natural boundary for both low and high temperature, we can only give (\ref{37}) as the conjectured singularities for the right arcs corresponding to $x>-1$ and $|k_{\mathrm{tr}}|=1$. Confirmation of this and a formula for the singularities on the left arcs can presumably be obtained by an analysis of the Vaidya (\cite{Vai}) integrals. 

\subsection{Fitting\label{sec:fit}}

Once all the singularities are suppressed, we fit the series to our scaling functions. We use only terms which are known (or assumed) to be nonzero, and leave out the known (and removed) Aharony and Fisher scaling terms. In other words, we fit to the linear combination
\begin{equation}\label{eq:fit}
(\sqrt{1 + \tau^2} + \tau)^{1/2} \tau^{-7/4} \left( a_6 \tau^6 + a_8 \tau^8 + a_{10} \tau^{10} + \ldots \right)
\end{equation}
with $a_6, a_8, a_{10}, \ldots$ our fitting coefficients.

Firstly, we convert each term in this expression from $\tau$ to our series variable $z$, as described in section \ref{sec:convert}. We then apply the singularity suppression that we applied to our susceptibility series to these fitting functions, so that the required equality between the two functions is maintained even though both functions have been changed by the suppression.

Finally we fit the suppressed series to the linear combination of our suppressed fitting functions. Suppose that the transformed and suppressed fitting function (\ref{eq:fit}) is $\sum_n f_n z^n$, while the subtracted and suppressed susceptibility series is $\sum_n c_n z^n$. We choose the amplitudes to minimise the expression
\begin{equation}
\sum_{n=n_1}^{n_2} \left( f_n - c_n \right)^2.
\end{equation}
The range of $n$ in the sum can be varied, but we always choose $n_2$ to be the largest available power of $z$ in our susceptibility series. In addition, varying $n_1$ will change the fitted amplitudes, which gives an idea of how accurate our fit is.

For the honeycomb lattice high-temperature series, we conduct two separate fits, one at the ferromagnetic point (with additional suppression of the antiferromagnetic singularity) and one at the antiferromagnetic point (with additional suppression of the ferromagnetic singularity). In fact we also did this for the square lattice, to check for an antiferromagnetic scaling term. We found no such scaling term, which is consistent with the results in \cite{ONGP}.

Once the initial fitting has been done, we can improve the accuracy of our fits by iteratively subtracting the new fit (or fits), re-suppressing singularities (replicating this in our fitting functions) and fitting again to the remainder, and so on.

\section{Acknowledgments}
AJG would like to thank Dr.\ Andrea Pelissetto and Dr.\ John Cardy for
patiently explaining aspects of conformal field theory as it applies to
the Ising model. We thank Dr.\ Iwan Jensen for making available to us
the 2000 term square-lattice susceptibility series that we have used in
this analysis. We also thank Dr.\ Barry McCoy for his interest.
This work was supported by the Australian Research Council
through a grant to MASCOS, the ARC Centre of Excellence for Mathematics and
Statistics of Complex Systems, thus supporting the work of YBC and AJG.
JHHP has been supported in part by the National Science Foundation under
grant PHY-07-58139 and by the Australian Research Council under Project
IDs LX0989627 and DP1096713.

\newpage

\appendix\section*{Appendices}

 \section{Ferromagnetic scaling function\label{app:ferro}}
 \subsection{Square lattice\label{app:ferrosq}}
\begin{eqnarray*}
F_-^{\mathrm{sq}}&=&(\tau + \sqrt{1+ \tau^2})^{\frac{1}{2}}(1+\tau^2/2-\tau^4/12\\
       &&-\;6.3213068404959366230670987124576163379333404464\backslash\\
      &&\hspace{16em}29429335850509012099708742399\cdot \tau^{6}\\
       &&+\;6.2519974704602432856837331806319562265626657486\backslash\\
      &&\hspace{16em}9581059930911004970341\cdot \tau^{8}\\
       &&-\;5.6896599756179940495694760341390552949459234168\backslash\\
      &&\hspace{16em}0072164185003897\cdot \tau^{10}\\
       &&+\;5.14221827114214604273511179366558788399868131986546472359\cdot \tau^{12}\\
       &&-\;4.67471611538219753943422533513538091798878146367647\cdot \tau^{14}\\
       &&+\;4.28351401741664147913747092020949150840022385\cdot \tau^{16}\\
       &&-\;3.93463085065515612248985707350481524149\cdot \tau^{18}\\
       &&+\;3.613033718221972872129117995447426\cdot \tau^{20}\\
       &&-\;3.3030941616500642890625665822\cdot \tau^{22}\\
       &&+\;2.99419136711436481655789\cdot \tau^{24}-2.674815242128336541\cdot \tau^{26}\\
       &&+\;2.3339198769874\cdot \tau^{28}-1.95837351\cdot \tau^{30}+1.537\cdot \tau^{32}),
\end{eqnarray*}

\begin{eqnarray*}
F_+^{\mathrm{sq}}&=&(\tau + \sqrt{1+ \tau^2})^{\frac{1}{2}}( 1+\tau^2/2-\tau^4/12 \\
       &&-\;0.12352922857520866639356466570562347322323268198504142433416176\cdot \tau^{6}\\
       &&+\;0.13661094980909643478343857458083310826834711524701276519\cdot \tau^{8}\\
       &&-\;0.13043897213329076084013583556244683622929916938362\cdot \tau^{10}\\
       &&+\;0.121512875791442694842447521021056149318718395\cdot \tau^{12}\\
       &&-\;0.1129603634344171840043033744010408654148\cdot \tau^{14}\\
       &&+\;0.10536961142693738687373469324338873\cdot \tau^{16}\\
       &&-\;0.0982140320131209895954107399728\cdot \tau^{18}\\
       &&+\;0.091314688764698386593329786\cdot \tau^{20}
         -0.08439419183682814997218\cdot \tau^{22}\\
       &&+\;0.0772604004964458205\cdot \tau^{24}-0.069668638313388\cdot \tau^{26}\\
       &&+\;0.061368727265\cdot \tau^{28}-0.05204288\cdot \tau^{30}+0.0414\cdot \tau^{32}).
       \end{eqnarray*}

\subsection{Triangular lattice\label{app:ferrotr}}
\begin{eqnarray*}
  F_-^{\mathrm{tr}}&=&(\tau + \sqrt{1+ \tau^2})^{\frac{1}{2}}(1+1/2\cdot \tau^2-21/256\cdot \tau^4\\
        &&-\;6.7764559898170749532861771919188746477857219070(3)\cdot \tau^{6}\\
        &&+\;6.84262914118601551543582352085826620764414(10)\cdot \tau^{8}\\
        &&-\;6.250933162702506214104998011755062095(9)\cdot \tau^{10}\\
        &&+\;5.63987692190321788346983658716286(30)\cdot \tau^{12}\\
        &&-\;5.106253322544511659092052061(5)\cdot \tau^{14}\\
        &&+\;4.65493974449161799368079(6)\cdot \tau^{16}\\
        &&-\;4.2701171199002454178(4)\cdot \tau^{18}\\
        &&+\;3.9327480363388237(23)\cdot \tau^{20}
          -3.625158242566(11)\cdot \tau^{22}\\
        &&+\;3.33306138(7)\cdot \tau^{24}-3.04765(11)\cdot \tau^{26}),
 \end{eqnarray*}
 
\begin{eqnarray*}
F_+^{\mathrm{tr}}&=&(\tau + \sqrt{1+ \tau^2})^{\frac{1}{2}}(1+1/2\cdot \tau^2-21/256\cdot \tau^4\\
       &&-\;0.1359799770448664282788192846845965785(4)\cdot \tau^{6}\\
       &&+\;0.152349558318015426490910429319733(17)\cdot \tau^{8}\\
       &&-\;0.14450411683821267150571729255(18)\cdot \tau^{10}\\
       &&+\;0.1331875171226390774852445(8)\cdot \tau^{12}\\
       &&-\;0.1223854265244510620558(16)\cdot \tau^{14}\\
       &&+\;0.1128620837499335229(18)\cdot \tau^{16}\\
       &&-\;0.1045232876841806(12)\cdot \tau^{18}\\
       &&+\;0.0970484952533(5)\cdot \tau^{20}
         -0.09008180554(18)\cdot \tau^{22}\\
       &&+\;0.08331757(8)\cdot \tau^{24}-0.07654(4)\cdot \tau^{26}).
\end{eqnarray*}

 \subsection{Honeycomb lattice\label{app:ferrohc}}
\begin{eqnarray*}
F_-^{\mathrm{hc}}&=&(\tau + \sqrt{1+ \tau^2})^{\frac{1}{2}}(1+1/2\cdot \tau^2-21/256\cdot \tau^4\\
        &&-\;2.2311493924390249844287257306396248825952406357(14)\cdot \tau^{6}\\
        &&+\;2.29732254380796554657837205957901644245366(35)\cdot \tau^{8}\\
        &&-\;2.169834272110400332644049880573751163(24)\cdot \tau^{10}\\
        &&+\;2.0232401262820557301956317745273(7)\cdot \tau^{12}\\
        &&-\;1.887361520807880372000774531(10)\cdot \tau^{14}\\
        &&+\;1.76703614250551894208334(10)\cdot \tau^{16}\\
        &&-\;1.6608531413942897306(6)\cdot \tau^{18}\\
        &&+\;1.5669217708308492(27)\cdot \tau^{20}
          -1.482989258248(11)\cdot \tau^{22}\\
        &&+\;1.40629074(6)\cdot \tau^{24}-1.33479(6)\cdot \tau^{26}),
\end{eqnarray*}

\begin{eqnarray*}
F_+^{\mathrm{hc}}&=&(\tau + \sqrt{1+ \tau^2})^{\frac{1}{2}}(1+1/2\cdot \tau^2-21/256\cdot \tau^4\\
       &&-\;0.01765738818162214275960642822819883(11)\cdot \tau^{6}\\
       &&+\;0.0340269694547711409716975728625(27)\cdot \tau^{8}\\
       &&-\;0.038525626277127202618271411(17)\cdot \tau^{10}\\
       &&+\;0.03947785106181932194262(4)\cdot \tau^{12}\\
       &&-\;0.03910559858848358918(5)\cdot \tau^{14}\\
       &&+\;0.03820301383607213(3)\cdot \tau^{16}\\
       &&-\;0.037081202839025(16)\cdot \tau^{18}\\
       &&+\;0.035888270323(6)\cdot \tau^{20}
         -0.0346847083(13)\cdot \tau^{22}\\
       &&+\;0.03347384(20)\cdot \tau^{24}-0.032213(27)\cdot \tau^{26}).
\end{eqnarray*}

\section{Antiferromagnetic scaling function\label{app:antiferro}}
\subsection{Honeycomb lattice\label{app:antiferrohc}}

\begin{eqnarray*}
F_-^{\mathrm{hc}}|^{\mathrm{af}}&=&
-\,(\tau+\sqrt{1+\tau^2})^{\frac{1}{2}}\\
&&\times\big(
4.545306597378049968857451461279249765190481271258(18)\cdot \tau^{6}\\
        &&\hspace{1.0em}-\;4.545306597378049968857451461279249765190481271258(18)\cdot \tau^{8}\\
        &&\hspace{1.0em}+\;4.0810988905921058814609481311813109325(5)\cdot \tau^{10}\\
        &&\hspace{1.0em}-\;3.61663679562116215327420481263559(4)\cdot \tau^{12}\\
        &&\hspace{1.0em}+\;3.2188918017366312870912775304(12)\cdot \tau^{14}\\
        &&\hspace{1.0em}-\;2.887903601986099051597450(19)\cdot \tau^{16}\\
        &&\hspace{1.0em}+\;2.60926397850595568720(21)\cdot \tau^{18}
          -2.3658262655079745(17)\cdot \tau^{20}\\
        &&\hspace{1.0em}+\;2.142168984318(13)\cdot \tau^{22}
          -1.92677064(12)\cdot \tau^{24}+1.7124(8)\cdot \tau^{26}\big),
\end{eqnarray*}
\begin{eqnarray*}
F_+^{\mathrm{hc}}|^{\mathrm{af}}&=&-\,(\tau+\sqrt{1+\tau^2})^{\frac{1}{2}}\\
&&\times\big(0.1183225888632442855192128564563977189(6)\cdot \tau^{6}\\
         &&\hspace{1.0em}-\;0.1183225888632442855192128564563977189(6)\cdot \tau^{8}\\
         &&\hspace{1.0em}+\;0.10597849056108546888744587531(10)\cdot \tau^{10}\\
         &&\hspace{1.0em}-\;0.0937096660608197555426090(10)\cdot \tau^{12}\\
         &&\hspace{1.0em}+\;0.083279827935967472857(4)\cdot \tau^{14}\\
         &&\hspace{1.0em}-\;0.074659069913861378(7)\cdot \tau^{16}
           +0.067442084845148(6)\cdot \tau^{18}\\
         &&\hspace{1.0em}-\;0.061160224928(3)\cdot \tau^{20}
           +0.0553970966(11)\cdot \tau^{22}\\
         &&\hspace{1.0em}-\;0.0498435(28)\cdot \tau^{24}+0.04424(7)\cdot \tau^{26}\big).
\end{eqnarray*}

Comparison of these results with what is obtained from the ferromagnetic expressions in Appendices \ref{app:ferrotr} and \ref{app:ferrohc} using (\ref{11}) yields a partial check of the consistency of our numerical fitting described in Section \ref{sec:extract}.

\section{Short-distance terms\label{app:short}}

Here we give the short-distance ``regular" background terms of the form
$$
B = \sum_{q=0}^\infty \sum_{p=0}^{\lfloor \sqrt{q} \rfloor} b^{(p,q)}(\log |\tau| )^p\tau^q
$$
rounded to 15 places. Our complete results are available in \cite{Sourcefiles}.

\subsection{Ferromagnetic square lattice\label{app:shortsq}}
{\small
\begin{eqnarray*}
B_{\mathrm{sq}}&=&(\tau+\sqrt{1+\tau^2})^{1/2}\\\nonumber
  &&\hspace{-1.4em}\times\;[-\,.104133245093831-.074368869753207\,\tau-.008144713909120\,\tau^2\\\nonumber
    &&+\,.004504107712232\,\tau^3+.239618794254722\,\tau^4-.002539950595339\,\tau^5\\\nonumber
    &&-\,.235288909669962\,\tau^6+.001915707531701\,\tau^7+.214340096611538\,\tau^8\\\nonumber
    &&-\,.000883215706003\,\tau^9-.194220628407196\,\tau^{10}+.000007233509777\,\tau^{11}\\\nonumber
    &&+\,.177102037555467\,\tau^{12}+.000688811096268\,\tau^{13}-.162792536489746\,\tau^{14}\\\nonumber
    &&-\,.001236572355315\,\tau^{15}+.150013412064378\,\tau^{16}+.001671694059110\,\tau^{17}\\\nonumber
    &&-\,.138208109106217\,\tau^{18}-.002022002972782\,\tau^{19}+.126799277310505\,\tau^{20}\\\nonumber
    &&+\,.002308285588780\,\tau^{21}-.115396441906289\,\tau^{22}-.002545765264414\,\tau^{23}\\\nonumber
    &&+\,.103574086263807\,\tau^{24}+.002745532102527\,\tau^{25}-.090922989554413\,\tau^{26}\\\nonumber
    &&-\,.002916073299270\,\tau^{27}+.076954225263348\,\tau^{28}+.003063568441388\,\tau^{29}\\\nonumber
 &&\hspace{-2.0em}+(\ln|\tau|)\\\nonumber
  &&\hspace{-1.4em}\times\;(+\,.032352268477309\,\tau-.005775529379688\,\tau^3+.059074961290345\,\tau^4\\\nonumber
    &&+\,.003058491575856\,\tau^5-.059166272208841\,\tau^6-.002067088393167\,\tau^7\\\nonumber
    &&+\,.054246930704214\,\tau^8+.001060102531550\,\tau^9-.049300253157083\,\tau^{10}\\\nonumber
    &&-\,.000268300641612\,\tau^{11}+.045027052571957\,\tau^{12}-.000343326832572\,\tau^{13}\\\nonumber
    &&-\,.041428586463053\,\tau^{14}+.000819393297118\,\tau^{15}+.038202673904453\,\tau^{16}\\\nonumber
    &&-\,.001196464684146\,\tau^{17}-.035217475800642\,\tau^{18}+.001500680711946\,\tau^{19}\\\nonumber
    &&+\,.032331741680806\,\tau^{20}-.001750700134389\,\tau^{21}-.029449221445927\,\tau^{22}\\\nonumber
    &&+\,.001959866123653\,\tau^{23}+.026464090269923\,\tau^{24}-.002137779981361\,\tau^{25}\\\nonumber
    &&-\,.023274239921560\,\tau^{26}+.002291702790868\,\tau^{27}+.019757464449312\,\tau^{28}\\\nonumber
    &&-\,.002426942629382\,\tau^{29})\\\nonumber
 &&\hspace{-2.0em}+(\ln|\tau|)^2\\\nonumber
  &&\hspace{-1.4em}\times\;(+\,.009391569871146\,\tau^4-.008695925462879\,\tau^6+.007669481493105\,\tau^8\\\nonumber
    &&+\,.000154284382979\,\tau^9-.006805407688144\,\tau^{10}-.000310520937481\,\tau^{11}\\\nonumber
\end{eqnarray*}
\begin{eqnarray*}
\phantom{B_{\mathrm{f}} }
    &&+\,.006113866432195\,\tau^{12}+.000444606198236\,\tau^{13}-.005557100215116\,\tau^{14}\\\nonumber
    &&-\,.000554418149346\,\tau^{15}+.005078042485427\,\tau^{16}+.000643607994970\,\tau^{17}\\\nonumber
    &&-\,.004649202184071\,\tau^{18}-.000716232782651\,\tau^{19}+.004246382079429\,\tau^{20}\\\nonumber
    &&+\,.000775832889819\,\tau^{21}-.003853404958387\,\tau^{22}-.000825213786325\,\tau^{23}\\\nonumber
    &&+\,.003454329481031\,\tau^{24}+.000866512510954\,\tau^{25}-.003034537504706\,\tau^{26}\\\nonumber
    &&-\,.000901440561715\,\tau^{27}+.002577451310655\,\tau^{28}+.000931250046525\,\tau^{29})\\\nonumber
 &&\hspace{-2.0em}+(\ln|\tau|)^3\\\nonumber
  &&\hspace{-1.4em}\times\;(-\,.000015771569138\,\tau^9+.000034428206621\,\tau^{11}-.000052442717749\,\tau^{13}\\\nonumber
    &&+\,.000068823835730\,\tau^{15}-.000002084325090\,\tau^{16}-.000083482363640\,\tau^{17}\\\nonumber
    &&+\,.000006458964601\,\tau^{18}+.000096589603855\,\tau^{19}-.000013639281329\,\tau^{20}\\\nonumber
    &&-\,.000108385585447\,\tau^{21}+.000023853448397\,\tau^{22}+.000119105615864\,\tau^{23}\\\nonumber
    &&-\,.000037600547029\,\tau^{24}-.000128947973257\,\tau^{25}+.000055460969100\,\tau^{26}\\\nonumber
    &&+\,.000138099034068\,\tau^{27}-.000078321412692\,\tau^{28}-.000146701272364\,\tau^{29})\\\nonumber
 &&\hspace{-2.0em}+(\ln|\tau|)^4\\\nonumber
  &&\hspace{-1.4em}\times\;(-\,.000000145427323\,\tau^{16}+.000000452982068\,\tau^{18}-.000000959267146\,\tau^{20}\\\nonumber
    &&+\,.000001683186013\,\tau^{22}-.000002660926741\,\tau^{24}-.000000003368087\,\tau^{25}\\\nonumber
    &&+\,.000003934622630\,\tau^{26}+.000000009693809\,\tau^{27}-.000005565949306\,\tau^{28}\\\nonumber
    &&-\,.000000023894457\,\tau^{29})\\\nonumber
 &&\hspace{-2.0em}+(\ln|\tau|)^5\\\nonumber
  &&\hspace{-1.4em}\times\;(+\,.000000000141953\,\tau^{25}-.000000000441519\,\tau^{27}+.000000001224727\,\tau^{29})].\\\nonumber
\end{eqnarray*}
}

\subsection{Antiferromagnetic square lattice\label{app:shortsqaf}}
{\small
\begin{eqnarray*}
B_{\mathrm{sq}}^{\mathrm{af}}&=&(\tau+\sqrt{1+\tau^2})^{1/2}\\\nonumber
  &&\hspace{-1.4em}\times\;[+\,.158866522960947+.149566836938536\,\tau+.010712225879833\,\tau^2\\\nonumber
    &&+\,.012753018839962\,\tau^3-.011741188869656\,\tau^4-.014066040875666\,\tau^5\\\nonumber
    &&+\,.013106454615626\,\tau^6+.012239696625538\,\tau^7-.011840194045411\,\tau^8\\\nonumber
    &&-\,.010585409302312\,\tau^9+.010151560037724\,\tau^{10}+.009080004112331\,\tau^{11}\\\nonumber
    &&-\,.008542012228790\,\tau^{12}-.007717026940132\,\tau^{13}+.007123677682511\,\tau^{14}\\\nonumber
    &&+\,.006508646391366\,\tau^{15}-.005912245104109\,\tau^{16}-.005467010793273\,\tau^{17}\\\nonumber
    &&+\,.004900133679335\,\tau^{18}+.004597213578999\,\tau^{19}-.004074131287647\,\tau^{20}\\\nonumber
    &&-\,.003893839411793\,\tau^{21}+.003417128190380\,\tau^{22}+.003339548120697\,\tau^{23}\\\nonumber
    &&-\,.002905973440848\,\tau^{24}-.002906261172134\,\tau^{25}+.002510579952576\,\tau^{26}\\\nonumber
    &&+\,.002559795034096\,\tau^{27}-.002197017191525\,\tau^{28}-.002268131101616\,\tau^{29}\\\nonumber
\end{eqnarray*}
\begin{eqnarray*}
\phantom{B_{\mathrm{af}} }
 &&\hspace{-2.0em}+(\ln|\tau|)\\\nonumber
  &&\hspace{-1.4em}\times\;(-\,.155317190158011\,\tau+.032067148145870\,\tau^3-.007716887572462\,\tau^4\\\nonumber
    &&-\,.015675211573817\,\tau^5-.000285542451537\,\tau^6+.009607254502732\,\tau^7\\\nonumber
    &&+\,.004835406420625\,\tau^8-.006064990344481\,\tau^9-.007340015041447\,\tau^{10}\\\nonumber
    &&+\,.003910356521404\,\tau^{11}+.008708427445682\,\tau^{12}-.002697783010885\,\tau^{13}\\\nonumber
    &&-\,.009405623038077\,\tau^{14}+.002161267525775\,\tau^{15}+.009683424894714\,\tau^{16}\\\nonumber
    &&-\,.002106502189570\,\tau^{17}-.009696425760611\,\tau^{18}+.002374179655585\,\tau^{19}\\\nonumber
    &&+\,.009556527066075\,\tau^{20}-.002827702235523\,\tau^{21}-.009355445823856\,\tau^{22}\\\nonumber
    &&+\,.003353500098021\,\tau^{23}+.009167728425385\,\tau^{24}-.003868501274293\,\tau^{25}\\\nonumber
    &&-\,.009041887308405\,\tau^{26}+.004329857870148\,\tau^{27}+.008988699114041\,\tau^{28}\\\nonumber
    &&-\,.004740651061453\,\tau^{29})\\\nonumber
 &&\hspace{-2.0em}+(\ln|\tau|)^2\\\nonumber
  &&\hspace{-1.4em}\times\;(+\,.011533714378823\,\tau^4-.011311734920692\,\tau^6+.010045768711199\,\tau^8\\\nonumber
    &&-\,.000475698571097\,\tau^9-.008783972022287\,\tau^{10}+.001157180172964\,\tau^{11}\\\nonumber
    &&+\,.007680651109513\,\tau^{12}-.001865091261620\,\tau^{13}-.006744701894515\,\tau^{14}\\\nonumber
    &&+\,.002491836298308\,\tau^{15}+.005964068368078\,\tau^{16}-.002972695839442\,\tau^{17}\\\nonumber
    &&-\,.005327647984236\,\tau^{18}+.003273731192324\,\tau^{19}+.004822634345908\,\tau^{20}\\\nonumber
    &&-\,.003388237983845\,\tau^{21}-.004425467442049\,\tau^{22}+.003335544654325\,\tau^{23}\\\nonumber
    &&+\,.004094631068058\,\tau^{24}-.003157926493823\,\tau^{25}-.003772517361799\,\tau^{26}\\\nonumber
    &&+\,.002912409430061\,\tau^{27}+.003402363586667\,\tau^{28}-.002655520057041\,\tau^{29})\\\nonumber
 &&\hspace{-2.0em}+(\ln|\tau|)^3\\\nonumber
  &&\hspace{-1.4em}\times\;(+\,.000057899719476\,\tau^9-.000169915088240\,\tau^{11}+.000326648846875\,\tau^{13}\\\nonumber
    &&-\,-.000517190858645\,\tau^{15}-.000001422188017\,\tau^{16}+.000729027463661\,\tau^{17}\\\nonumber
    &&+\,.000009170599968\,\tau^{18}-.000948102082432\,\tau^{19}-.000032108584334\,\tau^{20}\\\nonumber
    &&+\,.001159641637018\,\tau^{21}+.000080604994072\,\tau^{22}-.001349862803424\,\tau^{23}\\\nonumber
    &&-\,.000160502424863\,\tau^{24}+.001508463556122\,\tau^{25}+.000264448046627\,\tau^{26}\\\nonumber
    &&-\,.001631439353298\,\tau^{27}-.000364294406680\,\tau^{28}+.001722930696448\,\tau^{29})\\\nonumber
 &&\hspace{-2.0em}+(\ln|\tau|)^4\\\nonumber
  &&\hspace{-1.4em}\times\;(-\,.000000160856746\,\tau^{16}+.000000456983407\,\tau^{18}-.000000040918655\,\tau^{20}\\\nonumber
    &&-\,.000003417322708\,\tau^{22}+.000013997416073\,\tau^{24}-.000000009021984\,\tau^{25}\\\nonumber
    &&-\,.000036813182410\,\tau^{26}+.000000125169178\,\tau^{27}+.000075735555538\,\tau^{28}\\\nonumber
    &&-\,.000000779435552\,\tau^{29})\\\nonumber
 &&\hspace{-2.0em}+(\ln|\tau|)^5\\\nonumber
  &&\hspace{-1.4em}\times\;(-\,.000000001286222\,\tau^{25}+.000000004385050\,\tau^{27}+.000000027799861\,\tau^{29})].\\\nonumber
\end{eqnarray*}
}

\subsection{Ferromagnetic triangular lattice\label{app:shorttr}}
{\small
\begin{eqnarray*}
B_{\mathrm{tr}}&=&(\tau+\sqrt{1+\tau^2})^{1/2}\\\nonumber
  &&\hspace{-1.4em}\times\;[-\,.049561116521763-.029358763163227\,\tau-.003802085786368\,\tau^2\\\nonumber
    &&+\,.006390376143904\,\tau^3+.194331491416170\,\tau^4-.004659659320547\,\tau^5\\\nonumber
    &&-\,.195488838278358\,\tau^6+.003651173504528\,\tau^7+.178621656686715\,\tau^8\\\nonumber
    &&-\,.002895748949957\,\tau^9-.161336242614720\,\tau^{10}+.002311455809247\,\tau^{11}\\\nonumber
    &&+\,.146284971711630\,\tau^{12}-.001842503338574\,\tau^{13}-.133577942096403\,\tau^{14}\\\nonumber
    &&+\,.001456977882233\,\tau^{15}+.122753362974429\,\tau^{16}-.001134805649081\,\tau^{17}\\\nonumber
    &&-\,.113263489337451\,\tau^{18}+.000862240813452\,\tau^{19}+.104601807098273\,\tau^{20}\\\nonumber
    &&-\,.000629270080307\,\tau^{21}-.096359884476827\,\tau^{22}+.000428320304385\,\tau^{23}\\\nonumber
 &&\hspace{-2.0em}+(\ln|\tau|)\\\nonumber
  &&\hspace{-1.4em}\times\;(-\,.005374288589598\,\tau+.001021325616916\,\tau^3+.049253501657254\,\tau^4\\\nonumber
    &&-\,.000006005387528\,\tau^5-.050675128993180\,\tau^6-.000277768605459\,\tau^7\\\nonumber
    &&+\,.046680337431830\,\tau^8+.000300252836069\,\tau^9-.042334826787302\,\tau^{10}\\\nonumber
    &&-\,.000227268452048\,\tau^{11}+.038476859060257\,\tau^{12}+.000114673447726\,\tau^{13}\\\nonumber
    &&-\,.035187926212720\,\tau^{14}+.000011949942590\,\tau^{15}+.032370727904288\,\tau^{16}\\\nonumber
    &&-\,.000140047340708\,\tau^{17}-.029894415736149\,\tau^{18}+.000263535279625\,\tau^{19}\\\nonumber
    &&+\,.027634040948919\,\tau^{20}-.000379722086737\,\tau^{21}-.025487516010638\,\tau^{22}\\\nonumber
    &&+\,.000487692500440\,\tau^{23})\\\nonumber
 &&\hspace{-2.0em}+(\ln|\tau|)^2\\\nonumber
  &&\hspace{-1.4em}\times\;(+\,.008301571737990\,\tau^4-.007863822472801\,\tau^6+.006940825976817\,\tau^8\\\nonumber
    &&+\,.000004920887586\,\tau^9-.006124967722414\,\tau^{10}-.000008028657674\,\tau^{11}\\\nonumber
    &&+\,.005459215424842\,\tau^{12}+.000010897860945\,\tau^{13}-.004918734820641\,\tau^{14}\\\nonumber
    &&-\,.000014290481783\,\tau^{15}+.004471830270400\,\tau^{16}+.000018331038938\,\tau^{17}\\\nonumber
    &&-\,.004090720207698\,\tau^{18}-.000022918919968\,\tau^{19}+.003752754837032\,\tau^{20}\\\nonumber
    &&+\,.000027921194209\,\tau^{21}-.003440642618124\,\tau^{22}-.000033228452894\,\tau^{23})\\\nonumber
 &&\hspace{-2.0em}+(\ln|\tau|)^3\\\nonumber
  &&\hspace{-1.4em}\times\;(-\,.000008243826432\,\tau^9+.000019339918959\,\tau^{11}-.000030791103742\,\tau^{13}\\\nonumber
    &&+\,.000041558105858\,\tau^{15}-.000000289107374\,\tau^{16}-.000051264920936\,\tau^{17}\\\nonumber
    &&+\,.000001306093325\,\tau^{18}+.000059839322918\,\tau^{19}-.000003538519413\,\tau^{20}\\\nonumber
    &&-\,.000067344361243\,\tau^{21}+.000007394079439\,\tau^{22}+.000073895527830\,\tau^{23})\\\nonumber
 &&\hspace{-2.0em}+(\ln|\tau|)^4\\\nonumber
  &&\hspace{-1.4em}\times\;(-\,.000000023041822\,\tau^{16}+.000000103665399\,\tau^{18}-.000000280140679\,\tau^{20}\\\nonumber
    &&+\,.000000584762202\,\tau^{22})].\\\nonumber
\end{eqnarray*}
\vglue-1em\noindent
The leading term in $B_{\mathrm{tr}}$ above confirms the estimate in (26)
of \cite{MDMB} after adding a factor 2 needed because of a difference in
conventions.}

\subsection{Ferromagnetic honeycomb lattice\label{app:shorthc}}
{\small
\begin{eqnarray*}
B_{\mathrm{hc}}&=&(\tau+\sqrt{1+\tau^2})^{1/2}\\\nonumber
  &&\hspace{-1.4em}\times\;[-\,.221526277068482-.170518806873542\,\tau-.019236029093417\,\tau^2\\\nonumber
    &&-\,.000240258087320\,\tau^3+.140112831065240\,\tau^4+.002715126912573\,\tau^5\\\nonumber
    &&-\,.139853710977321\,\tau^6-.002422123613147\,\tau^7+.130966255841349\,\tau^8\\\nonumber
    &&+\,.002478441994620\,\tau^9-.121767984156181\,\tau^{10}-.002593394970413\,\tau^{11}\\\nonumber
    &&+\,.113507041635338\,\tau^{12}+.002697199234761\,\tau^{13}-.106296857456039\,\tau^{14}\\\nonumber
    &&-\,.002778848577189\,\tau^{15}+.099986481192228\,\tau^{16}+.002840749390802\,\tau^{17}\\\nonumber
    &&-\,.094433287150951\,\tau^{18}-.002887196461298\,\tau^{19}+.089489266542983\,\tau^{20}\\\nonumber
    &&+\,.002921915790808\,\tau^{21}-.084984336381365\,\tau^{22}-.002947776518703\,\tau^{23}\\\nonumber
 &&\hspace{-2.0em}+(\ln|\tau|)\\\nonumber
  &&\hspace{-1.4em}\times\;(+\,.110304596706594\,\tau-.017367191250168\,\tau^3+.032554394731493\,\tau^4\\\nonumber
    &&+\,.007749610093406\,\tau^5-.033370773266168\,\tau^6-.004545306065368\,\tau^7\\\nonumber
    &&+\,.031517394252347\,\tau^8+.002697183732340\,\tau^9-.029438657077664\,\tau^{10}\\\nonumber
    &&-\,.001512880972029\,\tau^{11}+.027527821013048\,\tau^{12}+.000704802950233\,\tau^{13}\\\nonumber
    &&-\,.025841562449391\,\tau^{14}-.000126907282155\,\tau^{15}+.024355378143637\,\tau^{16}\\\nonumber
    &&-\,.000302184134792\,\tau^{17}-.023041033316250\,\tau^{18}+.000630617459438\,\tau^{19}\\\nonumber
    &&+\,.021866143676416\,\tau^{20}-.000888340903130\,\tau^{21}-.020791271125817\,\tau^{22}\\\nonumber
    &&+\,.001094822052575\,\tau^{23})\\\nonumber
 &&\hspace{-2.0em}+(\ln|\tau|)^2\\\nonumber
  &&\hspace{-1.4em}\times\;(+\,.004328421950579\,\tau^4-.004173174221495\,\tau^6+.003864856815018\,\tau^8\\\nonumber
    &&+\,.000098594831882\,\tau^9-.003581293629898\,\tau^{10}-.000203109598421\,\tau^{11}\\\nonumber
    &&+\,.003336710555165\,\tau^{12}+.000293321457772\,\tau^{13}-.003126961910309\,\tau^{14}\\\nonumber
    &&-\,.000367344803112\,\tau^{15}+.002944831872355\,\tau^{16}+.000427433155150\,\tau^{17}\\\nonumber
    &&-\,.002785081058367\,\tau^{18}-.000476286169756\,\tau^{19}+.002642924163304\,\tau^{20}\\\nonumber
    &&+\,.000516221811618\,\tau^{21}-.002513107496440\,\tau^{22}-.000549081268363\,\tau^{23})\\\nonumber
 &&\hspace{-2.0em}+(\ln|\tau|)^3\\\nonumber
  &&\hspace{-1.4em}\times\;(-\,.000008459084030\,\tau^9+.000018328291997\,\tau^{11}-.000027731894823\,\tau^{13}\\\nonumber
    &&+\,.000036311733546\,\tau^{15}-.000000100196269\,\tau^{16}-.000044084313016\,\tau^{17}\\\nonumber
    &&+\,.000000291941736\,\tau^{18}+.000051136242275\,\tau^{19}-.000000616104491\,\tau^{20}\\\nonumber
    &&-\,.000057556878974\,\tau^{21}+.000001226160802\,\tau^{22}+.000063427686583\,\tau^{23})\\\nonumber
 &&\hspace{-2.0em}+(\ln|\tau|)^4\\\nonumber
  &&\hspace{-1.4em}\times\;(-\,.000000008316207\,\tau^{16}+.000000024195538\,\tau^{18}-.000000049128873\,\tau^{20}\\\nonumber
    &&+\,.000000092415863\,\tau^{22})].\\\nonumber
\end{eqnarray*}
}

\subsection{Antiferromagnetic honeycomb lattice\label{app:shorthcaf}}
{\small
\begin{eqnarray*}
B_{\mathrm{hc}}^{\mathrm{af}}&=&(\tau+\sqrt{1+\tau^2})^{1/2}\,\big[\,.122404044024957+.111801280547087\,\tau+\ldots\\\nonumber
 &&\hspace{1.0em}+(\ln|\tau|)\,(-\,.121053173885789\,\tau+\ldots)+\ldots\big],\\\nonumber
\end{eqnarray*}
}
as given more fully by (\ref{12}) and Appendices \ref{app:shorttr} and \ref{app:shorthc}.


\begin{thebibliography}{99}

\bibitem{Domb1}{%
C.\ Domb,
Order-Disorder Statistics.\ II. A Two-Dimensional Model,
{\it Proc.\ R.\ Soc.\ Lond.\ A} {\bf 199} (1949) 199--221.}

\bibitem{Domb2}{%
Series Expansions for Lattice Models,
{\it Phase Transitions and Critical Phenomena}, Vol. 3, 
C.\ Domb and M.S.\ Green, eds., Academic Press, London (1974).}

\bibitem{ONGP2}{%
W.P.\ Orrick, B.G.\ Nickel, A.J.\ Guttmann, and J.H.H.\ Perk,
Critical behavior of the two-dimensional Ising susceptibility,
{\it Phys.\ Rev.\ Lett.}\ {\bf 86} (2001) 4120--4123.
See also arXiv:cond-mat/0009059.}

\bibitem{ONGP}{%
W.P.\ Orrick, B.\ Nickel, A.J.\ Guttmann, and J.H.H.\ Perk,
The susceptibility of the square lattice Ising model: New developments,
{\it J.\ Stat.\ Phys.} {\bf 102} (2001) 795--841.
For the complete set of series coefficients, see http://www.ms.unimelb.edu.au/\~{}tonyg.
Preprint at arXiv:cond-mat/0103074.}

\bibitem{Nic1}
B.\ Nickel,
On the singularity structure of the 2D Ising model susceptibility,
{\it J.\ Phys.\ A: Math.\ Gen.} {\bf 32} (1999) 3889--3906.

\bibitem{Wu76}
T.T.\ Wu, B.M.\ McCoy, C.A.\ Tracy, and E.\ Barouch,
Spin-spin correlation functions for the two dimensional Ising model:
exact theory in the scaling region,
{\it Phys.\ Rev.\ B} {\bf 13} (1976) 315--374.

\bibitem{BMW73}
E.\ Barouch, B.M.\ McCoy, and T.T.\ Wu,
Zero-Field Susceptibility of the Two-Dimensional Ising Model near $T_c$,
{\it Phys.\ Rev.\ Lett.} {\bf 31} (1973) 1409--1411.

\bibitem{TM73}
C.A.\ Tracy and B.M.\ McCoy,
Neutron Scattering and the Correlation Functions of the Two-Dimensional Ising Model near $T_c$,
{\it Phys.\ Rev.\ Lett.} {\bf 31} (1973) 1500--1504.

\bibitem{JM17}{%
M.\ Jimbo and T.\ Miwa,
Studies on holonomic quantum fields. XVII,
{\it Proc.\ Japan Acad.\ A} {\bf 56} (1980) 405--410.
Errata {\bf 57} (1987) 347.}

\bibitem{AP2}{%
H.\ Au-Yang and J.H.H.\ Perk,
Correlation Functions and Susceptibility in the $Z$-Invariant Ising Model,
in {\it MathPhys Odyssey 2001: Integrable Models and Beyond,}
M.~Kashiwara and T.~Miwa, eds., (Birkh\"auser, Boston, (2002), pp.\ 23--48.
\hfill\break
Preprint at http://physics.okstate.edu/perk/papers/kyoto/ziising.pdf$\,$.}

\bibitem{Gu74}
A.J.\ Guttmann,
Susceptibility amplitudes for the two-dimensional Ising model,
{\it Phys.\ Rev.\ B} {\bf 9} (1974) 4991--4992,
{\it B} {\bf 12} (1975) 1991.

\bibitem{RB75}
D.S.\ Ritchie and D.D.\ Betts,
Extended universality of the Ising model,
{\it Phys.\ Rev.\ B} {\bf 11} (1975) 2559--2563.

\bibitem{AP03}
H.\ Au-Yang and J.H.H.\ Perk,
Susceptibility calculations in periodic and
quasiperiodic planar Ising models,
{\it Physica A} {\bf 321} (2003) 81--89.

\bibitem{Kong}
X.P.\ Kong,
Wave Vector Dependent Susceptibility of the
Two Dimensional Ising Model,
Ph.D.\ Thesis, State University of New York at Stony Brook (1987).

\bibitem{Fisher}
M.E.\ Fisher,
Transformations of Ising models,
{\it Phys.\ Rev.} {\bf 113} (1959) 969--981.

\bibitem{AF83}
A.\ Aharony and M.E.\ Fisher,
Nonlinear scaling fields and corrections to scaling near criticality,
\newblock{{\it Phys.\ Rev.\ B}  {\bf 27} (1983) 4394--4400.}

\bibitem{AF80}
A.\ Aharony and M.E.\ Fisher,
Universality in analytic corrections to scaling for planar Ising models,
\newblock{{\it Phys.\ Rev.\ Lett.}  {\bf 45} (1980) 679--682.}

\bibitem{Gut}
A.J.\ Guttmann,
Ising model amplitudes and extended lattice-lattice scaling,
{\it J. Phys. A: Math. Gen.} {\bf 10} (1977) 1911--1916.

\bibitem{GG78}
D.S.\ Gaunt and A.J.\ Guttmann,
A generalised form of extended lattice-lattice scaling,
{\it J.\ Phys.\ A: Math.\ Gen.} {\bf 11} (1978) 1381--1397.

\bibitem{MDMB}
V.V.\ Mangazeev, M.Yu.\ Dudalev, V.V.\ Bazhanov, and M.T.\ Batchelor,
Scaling and universality in the two-dimensional Ising model
with a magnetic field,
{\it Phys.\ Rev.\ E} {\bf 81} (2010) 060103(R).
Preprint at arXiv:1002.4234.

\bibitem{C02}
M.\ Caselle, M.\ Hasenbusch, A.\ Pelissetto and E.\ Vicari,
Irrelevant operators in the two-dimensional Ising model,
{\it J.\ Phys.\ A: Math.\ Theor.} {\bf 35} (2002) 4861--4888.
Preprint at arXiv:cond-mat/0106372.

\bibitem{GR}
I.S.\ Gradshteyn and I.M.\ Ryzhik,
Table of Integrals, Series, and Products,
fourth edition, Academic Press, New York (1980).

\bibitem{Syo}
I.\ Syozi, Transformation of Ising Models.
In C.\ Domb and M.S.\ Green (eds.),
{\it Phase transitions and critical phenomena, volume 1: exact results},
(1972), Academic Press, London.

\bibitem{P}{%
J.H.H.\ Perk,
Quadratic identities for Ising model correlations,
{\it Phys.\ Lett.\ A} {\bf 79} (1980) 3--5.}

\bibitem{MWa}{%
B.M.\ McCoy and T.T.\ Wu,
Nonlinear Partial Difference Equations
for the Two-Dimensional Ising Model,
{\it Phys.\ Rev.\ Lett.} {\bf 45} (1980) 675--678.}

\bibitem{MWb}{%
B.M.\ McCoy and T.T.\ Wu,
Nonlinear Partial Difference Equations for the Two-Spin
Correlation Function of the Two-Dimensional Ising Model,
{\it Nucl.\ Phys.\ B} {\bf 180[FS2]} (1980) 89--115.}

\bibitem{MPW}{%
B.M.\ McCoy, J.H.H.\ Perk, and T.T.\ Wu,
Ising Field Theory: Quadratic Difference Equations for
the $n$-Point Green's Functions on the Square Lattice,
{\it Phys.\ Rev.\ Lett.} {\bf 46} (1981) 757--760.}

\bibitem{MWbook}{%
B.M.\ McCoy and T.T.\ Wu,
The Two-Dimensional Ising Model,
Harvard University Press, Cambridge, Massachusetts (1973).}

\bibitem{B}{%
R.J.\ Baxter,
Solvable eight-vertex model on an arbitrary planar lattice,
{\it Phil.\ Trans.\ R.\ Soc.\ Lond.\ A} {\bf 289} (1978) 315--346.}

\bibitem{APZI}{%
H.\ Au-Yang and J.H.H.\ Perk,
Critical correlations in a $Z$-invariant inhomogeneous Ising model,
{\it Physica A} {\bf 144} (1987) 44--104.}

\bibitem{AP1}{%
H.\ Au-Yang and J.H.H.\ Perk,
Wavevector-Dependent Susceptibility in Aperiodic Planar Ising Models,
in {\it MathPhys Odyssey 2001: Integrable Models and Beyond,}
M.~Kashiwara and T.~Miwa, eds., (Birkh\"auser, Boston, (2002), pp.\ 1--21.
\hfill\break
Preprint at http://physics.okstate.edu/perk/papers/kyoto/triquasi.pdf$\,$.}

\bibitem{Naya}{%
S.\ Naya,
On the Spontaneous Magnetizations of Honeycomb and Kagom\'e Ising Lattices,
{\it Progr.\ Theor.\ Phys.}  {\bf 11} (1954) 53--62.}

\bibitem{W}{%
N.S.\ Witte,
Isomonodromic deformation theory and the next-to-diagonal
correlations of the anisotropic square lattice Ising model,
{\it J.\ Phys.\ A: Math.\ Theor.} {\bf 40} (2007) F491--F501.
Preprint at arXiv:0705.0557.}

\bibitem{H}{%
R.M.F.\ Houtappel,
``Order-Disorder in Hexagonal Lattices,"
Physica {\bf 16} (1950) 425--455.}

\bibitem{Wan}{%
G.H.\ Wannier,
``Antiferromagnetism. The Triangular Ising Net,"
Phys.\ Rev.\ {\bf 79} (1950) 357--364;
``Errata," Phys.\ Rev.\ B {\bf 7} (1973) 5017.}

\bibitem{N}{%
G.F.\ Newell,
``Crystal Statistics of a Two-Dimensional Triangular Ising Lattice,"
Phys.\ Rev.\ {\bf 79} (1950) 876--882.}

\bibitem{Ha}{%
H.\ Hancock,
``Lectures on the Theory of Elliptic Functions,"
Dover Publ., New York (1958), Art.\ 251, 357--360.}

\bibitem{WW}{%
E.T.\ Whittaker and G.N.\ Watson,
``A Course of Modern Analysis,"
fourth ed., Cambridge Univ.\ Press, Cambridge, U.K.\ (1927),
Ch.\ 21 and 22.}

\bibitem{Stephenson}
J.\ Stephenson,
Ising model spin correlations on the triangular lattice,
{\it J.\ Math.\ Phys.} {\bf 5} (1964) 1009--1024.

\bibitem{Sykes1}
M.F.\ Sykes, D.S.\ Gaunt, J.L.\ Martin, S.R.\ Mattingly, and J.W.\ Essam,
Derivation of low-temperature expansions for Ising model. IV.
Two-dimensional lattices: temperature grouping,
{\it J.\ Math.\ Phys.} {\bf 14} (1973) 1071--1074.

\bibitem{Sykes2}
M.F.\ Sykes, D.S.\ Gaunt, P.D.\ Roberts, and J.A.\ Wyles,
High temperature series for the susceptibility of the Ising model. I. Two dimensional lattices,
{\it J.\ Phys.\ A} {\bf 5} (1972) 624--639.

\bibitem{Sykes3}
M.F.\ Sykes, M.G.\ Watts, and D.S.\ Gaunt,
Derivation of low-temperature expansions for Ising model. VIII.
Ferromagnetic and antiferromagnetic polynomials for the honeycomb-triangular system,
{\it J.\ Phys.\ A: Math.\ Gen.} {\bf 8} (1975) 1448--1460.

\bibitem{Sourcefiles}
Y.\ Chan, A.J.\ Guttmann, B.G.\ Nickel, and J.H.H.\ Perk,
Additional material added to the source files of arXiv:1012.5272.

\bibitem{Bou}
S.\ Boukraa, A.J.\ Guttmann, S.\ Hassani, I.\ Jensen, J.-M.\ Maillard,
B.\ Nickel, and N.\ Zenine,
Experimental mathematics on the magnetic susceptibility of the square lattice Ising model,
{\it J.\ Phys.\ A: Math.\ Theor.} {\bf 41} (2008) 455202.
See http://www.ms.unimelb.edu.au/$\sim$iwan/ising/Ising\_ser.html
for the series coefficients.

\bibitem{Nic2}
B.\ Nickel,
Addendum to `On the singularity structure of the 2D ising model susceptibility,'
{\it J.\ Phys A: Math.\ Gen.} {\bf 33} (2000) 1693--1711.

\bibitem{Vai}
H.G.\ Vaidya,
The spin-spin correlation functions and susceptibility amplitudes
for the two-dimensional Ising model: triangular lattice.
{\it Phys.\ Lett.\ A} {\bf 57} (1976) 1--4.

\bibitem{Mat}
V.\ Matveev and R.\ Shrock,
Complex-temperature singularities in the $d=2$ Ising model: Triangular and honeycomb lattices,
{\it J.\ Phys A: Math.\ Gen.} {\bf 29} (1996) 803--823.
Preprints at arXiv:hep-lat/9411023 and arXiv:hep-lat/9412076.

\end{thebibliography}
\end{document}